\documentclass[usenatbib]{mn2e}
\usepackage{epsfig}
\usepackage{times}

\title[The planetary nebula NGC\,1501] {Observations and 3D photoionisation modelling of the Wolf-Rayet planetary nebula NGC\,1501}\author[Ercolano et al.]{B. Ercolano$^1$, R. Wesson$^1$, Y. Zhang$^2$, M.J. Barlow$^1$, O. De Marco$^3$, T. Rauch$^{4,5}$, X.-W. Liu$^2$\\
$^1$Department of Physics and Astronomy, University College London, Gower St, London WC1E~6BT, UK\\
$^2$Department of Astronomy, Peking University, Beijing 100871, P. R. China\\
$^3$Department of Astrophysics, American Museum of Natural History, Central Park West at 79th Street, NY 10024, USA\\
$^4$Institut f\"ur Astronomie und Astrophysik, Abt. Astronomie, Sand 1, Universit\"at T\"ubingen, 72076 T\"ubingen, Germany\\
$^5$Dr.-Remeis-Sternwarte, Sternwartstra\ss e 7, D-96049 Bamberg, Germany\\}
\date{Received:}

\begin{document}
\maketitle

\begin{abstract}

Deep optical spectra of the high excitation planetary nebula NGC~1501 and its W04 central star are presented. A recombination line abundance analysis of the central star's emission-line spectrum yields He:C:O mass fractions of 0.36:0.48:0.16, similar to those of PG1159 stars. A detailed empirical analysis of the nebular collisionally excited line (CEL) and optical recombination line (ORL) spectrum is presented, together with fully three-dimensional photoionisation modelling of the nebula. We found very large ORL-CEL abundance discrepancy factors (ADFs) for O$^{2+}$ (32) and Ne$^{2+}$ (33). The mean value of $\sim$5100~K for the  $T_{\rm e}$ derived from He~{\sc i} recombination lines ratios is 6000~K lower than the value of 11100~K implied by the [O~{\sc iii}] line ratio. This result indicates the existence of a second, low-temperature nebular component which could account for the observed ORL emission. Electron temperature fluctuations ($t^2$) cannot account for the high ADFs found from our optical spectra of this nebula. 

A three-dimensional photoionisation model of NGC~1501 was constructed using the photoionisation code \mbox{{\sc mocassin}}, based on our new spectroscopic data and using the three-dimensional electron density distribution determined from long-slit echellograms of the nebula by Ragazzoni et al. (2001). The central star ionising radiation field is approximated by a model atmosphere, calculated using the T\"ubingen NLTE Model Atmosphere Package \citep{rauch03}, for abundances typical of the W04 nucleus of NGC~1501 and PG1159 stars. The nebular emission line spectrum was best reproduced using a central star model with effective temperature T$_{\rm eff}$~=~110\,kK and luminosity $L_{\rm *}$~=~5000\,L$_{\odot}$. The initial models showed higher degrees of ionisation of heavy elements than indicated by observations. We investigated the importance of the missing low-temperature dielectronic recombination rates for third-row elements and have estimated upper limits to their rate coefficients.

Our single-phase, three-dimensional photoionisation model heavily under-predicts the optical recombination line emission. We conclude that the presence of a hydrogen-deficient, metal-rich component is necessary to explain the observed ORL spectrum of this object. The existence of such knots could also provide a softening of the radiation field, via the removal of ionising photons by absorption in the knots, thereby helping to alleviate the over-ionisation of the heavy elements in our models.

\end{abstract}

\nokeywords


\section{Introduction}
The high excitation, northern planetary nebula (PN) NGC 1501 has been the subject of several recent tomographic and spatio-kinematical studies \citep{sabbadin82, ragazzoni01, sabbadin00a}. The central star of NGC~1501 has been classified as OVI/WC4/WO4 \citep{aller76,tylenda93,gorny95,crowther98}, and has also been the subject of several studies, particularly due to it being one of the few known pulsating PG1159-type PN nuclei \citep{bond93, bond96, ciardullo96}. A detailed study of the ionised nebular gas, has however, not been carried out to date, due to the lack of suitable spectroscopic observations. In this paper we present a spectroscopic analysis and three-dimensional photoionisation modelling of the ionised gas in NGC 1501. Based on new optical observations and using the three-dimensional photoionisation code \mbox{{\sc mocassin}} \citep{ercolano03I} and the density distribution derived by Ragazzoni et al. (2001), we construct models in order to obtain better constraints on the nebular physical properties and elemental abundances. 

 It is a well known and widely studied problem that optical recombination lines (ORLs) in gaseous nebulae yield higher empirical abundances than collisionally excited lines \citep[CELs; for a recent review see ][]{liu02}; this is tightly correlated with Balmer jump (BJ) temperatures, T$_{\rm e}$(BJ), being systematically lower than those from the [O~{\sc iii}] nebular to auroral line ratio \citep{liu93}. One possible solution to the problem is that temperature and density fluctuations within the nebular gas lead to underestimated elemental abundances from standard CEL analyses \citep{peimbert67, peimbert71}. While this may be a valid explanation for objects with low abundance discrepancy factors \citep[e.g.][]{peimbert04}, for cases where T$_{\rm e}$(BJ) is more than a factor of 2 lower than T$_{\rm e}$([O~{\sc iii}]) the method is no longer applicable \citep{liu02}. Moreover, the $t^2$ values derived from T$_{\rm e}$(BJ) and T$_{\rm e}$([O~{\sc iii}]) for some less extreme nebulae are found to be too small to reconcile the ORL and CEL abundances \citep{liu02}. The main problem encountered with the $t^2$ formulation is the fact that the $T_{\sc e}$-insensitive IR fine-structure lines also yield much lower ionic abundances than those implied by ORLs \citep[e.g.][]{rubin97,liu00, liu01a, liu01b,tsamis04}. The presence of hydrogen-deficient metal-rich knots of high density embedded in the normal nebular gas has been proposed as an alternative to temperature fluctuations for the case of the PN~NGC~6153 \citep{liu00}. Dual abundance photoionisation models have been constructed for PNe NGC~6153 \citep{pequignot03, tylenda03}, M1-42 and M2-36 \citep{tylenda03} and for the hydrogen deficient knots of Abell~30 \citep{ercolano03III}. As discussed later in this paper, temperature and density fluctuations in a chemically homogeneous medium are not sufficient to explain the ionic abundance discrepancy factors (ADF's) derived from our observations of NGC~1501. 

The new observational data is presented in Section~2, followed by a detailed empirical analysis of the ORL and CEL data, which is presented in Section~3. In that section we also carry out a study of the central star spectrum, providing identifications and measured equivalent widths of the detected stellar emission lines, as well as an abundance analysis for the central star wind. The three-dimensional photoionisation modelling procedures are set out in Section~4, while Section~5 is concerned with the results of the modelling. After an investigation of the applicability of the $t^2$ formalism to our results, carried out in Section~\ref{sec:knots}, there follows a discussion of how the presence of hydrogen-deficient, metal-rich knots could resolve the remaining discrepancies between ORL and CEL abundances as well as the over-ionisation of the nebular gas as predicted by the modelling. Our final conclusions are given in Section~7 of the paper.

\begin{figure*}
\vspace{0.4cm}
 \centering \epsfig{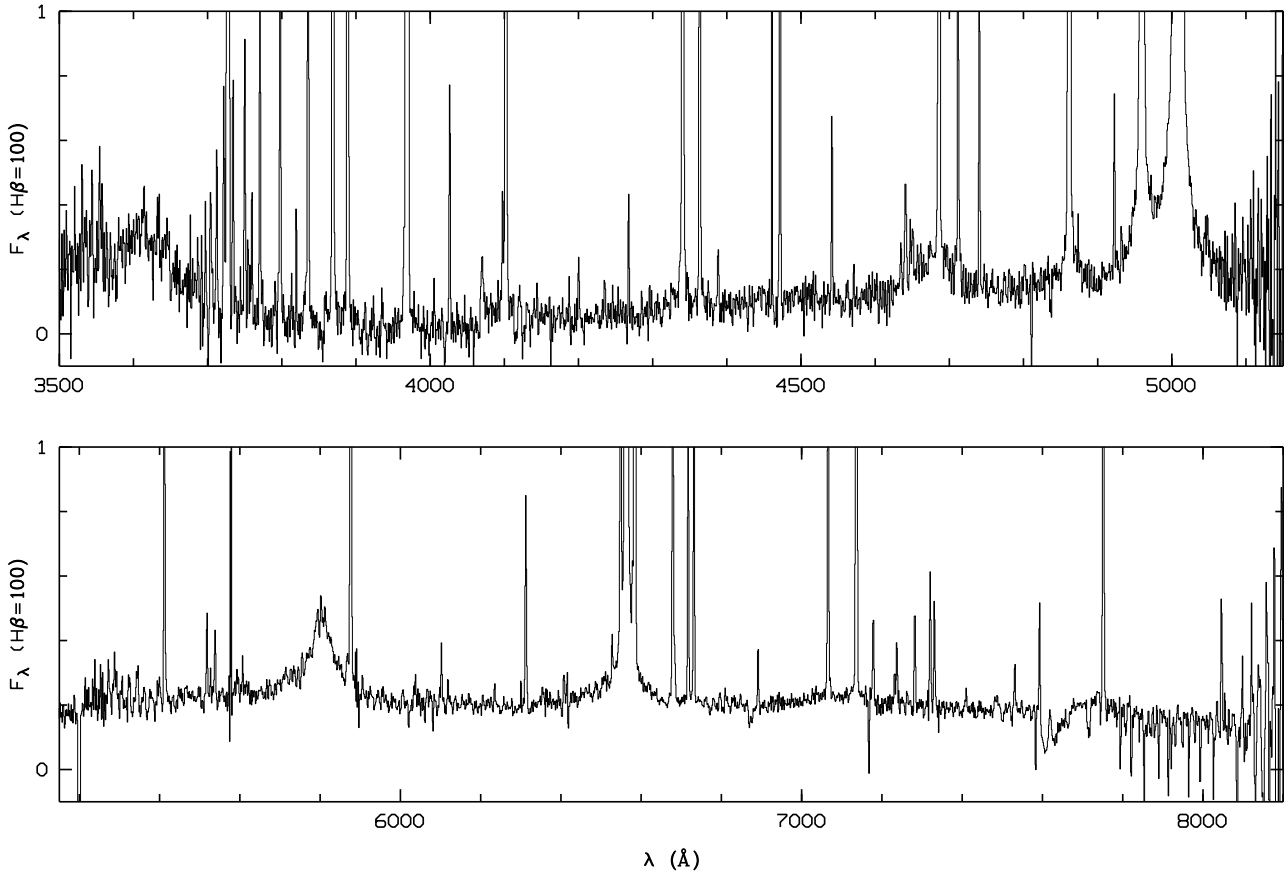}
\caption{The observed optical spectrum of the PN~NGC~1501 integrated over the angular extent of the nebula (68 arcsec) and normalised such that F(H$\beta$)~=~100.}

\label{spectrum}
\end{figure*}

\section{Observations}

In this section new observations of the optical spectrum of NGC~1501 are presented. The nebular analysis and the three-dimensional modelling which follow are all based on the new data. IUE spectra for the PN central star were also retrieved from the archive to provide some indication of the stellar energy distribution. These are also briefly discussed in this section.

\subsection{Optical Observations of NGC 1501}
\label{sub:optical}

The observations were carried out in dark sky conditions on the night of 2003 August 1st, using the double-armed ISIS long slit spectrograph mounted on the 4.2~m William Herschel Telescope at the Observatorio del Roque de los Muchachos, on La Palma, Spain.  The gratings used were the R316R in the red and the R600B in the blue, giving wavelength coverage from 3400-5100\,{\AA} in the blue and 5100-8000\,{\AA} in the red.  The spectrograph slit was aligned North-South, and the slit width used was 0.75\,arcsec (the slit length was 4\,arcmin), giving a spectral resolution of 1.5~{\AA} FWHM in the blue and 2.8~{\AA} FWHM in the red.  Five exposure were taken, of 1800~s each.

The data were reduced using standard {\sc LONG92} procedures in {\sc MIDAS}.  The 2D frames were bias-subtracted, flat-fielded, and cleaned of cosmic rays.  They were wavelength-calibrated using exposures of a CuNe+CuAr calibration lamp.  Spectra were flux-calibrated using wide slit (8\arcsec) observations of the standard stars HZ44 and Feige 110. Unfortunately wide slit observations of NGC~1501 were not acquired and, as a result, absolute flux calibration of its central star could not be carried out.

The top and bottom panels of Figure~\ref{spectrum} show the extracted blue and red spectra of NGC~1501 after integration over $\sim$68\,arcsec (338~pixels) along the slit. 

\subsection{IUE Observations of NGC 1501}
\label{sub:IUE}

Low- and high-dispersion ultra-violet spectra are available for NGC~1501 from the IUE archive. These have already been described by \citet{feibelman98}. Only the low-dispersion data is useful as the high-dispersion spectrum seems to be very underexposed. Moreover, as noticed by \citet{feibelman98}, the nebular contribution to the IUE data is very small, as indicated by the weakness of the C~{\sc iii}]~$\lambda\lambda$1906,1908 emission doublet, which is normally one of the major nebular emission features. Therefore, the IUE spectra could not be used for the nebular analysis carried out in this work. Nevertheless, the retrieved IUE SWP28953 and LWP08948 spectra were useful in providing some constraint on the luminosity of the central star, as described in Section~\ref{sub:cspar}.

\section{Empirical Analysis}

\subsection{The Nebular Spectrum}
A full list of observed lines and their measured fluxes is presented in Table~\ref{tab:linelist}. All line fluxes, except those of the strongest lines, were measured using Gaussian line profile fitting. The fluxes of the strongest isolated lines were obtained by integration over the detected profile.
 Column 1 of Table~\ref{tab:linelist} gives the observed wavelengths corrected for the heliocentric radial velocity of the nebula as determined from the Balmer lines. The observed fluxes are given in column 2. Column 3 lists the fluxes after correction for interstellar extinction, according to $I(\lambda)=10^{c({\rm H}\beta)f(\lambda)}F(\lambda)$, where $f(\lambda)$ is the standard Galactic extinction law for a total-to-selective extinction ratio of $R=3.1$ \citep{howarth83} and c(H$\beta$) is the logarithmic extinction at ${\rm H}\beta$ (see Section~\ref{sub:reddening}). The ionic identification, laboratory wavelength, multiplet number, the lower and upper spectral terms of the transition, and the statistical weights of the lower and upper levels, are given in columns 4--10, respectively. All fluxes are given relative to  ${\rm H}\beta$, on a scale where ${\rm H}\beta = 100$.

 \begin{table*}
 \tabcolsep 10pt
 \centering
 \caption{\label{tab:linelist} Observed and dereddened relative line fluxes, on a scale where H$\beta=100$. The integrated observed H($\beta$) flux from Cahn et al. (1992) was dereddened using c(H$\beta$)\,=\,1.0 to give an integrated dereddened flux of $I$(H$\beta$)~=~5.26\,$\cdot$\,10$^{-11}$\,erg\,s$^{-1}$\,cm$^{-2}$}.
 \begin{tabular}{lccccccccc}
 \hline
 \hline
 \noalign{\smallskip}
 $\lambda_{\rm obs}$&$F(\lambda)$&                  $I(\lambda)$&Ion&$\lambda_{\rm lab}$                           &Mult&Lower Term&Upper Term&$g_1$&$g_2$\\
 (1) & (2) & (3) & (4) & (5) & (6)                     & (7) & (8)& (9)& (10) \\
 \noalign{\smallskip}
 \hline
 \noalign{\smallskip}
 3614.02&     0.56&      1.07&  He I    &  3613.64& V6     &   2s  1S  &  5p  1P*&  1&  3\\
 3686.90&     0.39&      0.71&  H 19    &  3686.83& H19    &   2p+ 2P* & 19d+ 2D &  8&  *\\
 3691.80&     0.34&      0.63&  H 18    &  3691.56& H18    &   2p+ 2P* & 18d+ 2D &  8&  *\\
 3697.14&     0.74&      1.35&  H 17    &  3697.15& H17    &   2p+ 2P* & 17d+ 2D &  8&  *\\
 3703.88&     1.15&      2.09&  H 16    &  3703.86& H16    &   2p+ 2P* & 16d+ 2D &  8&  *\\
 3706.55&     0.45&      0.83&  He I    &  3705.02& V25    &   2p  3P* &  7d  3D &  9& 15\\
       &         *&         *&  O III   &  3707.25& V14    &   3p  3P  &  3d  3D*&  3&  5\\
 3712.21&     1.43&      2.60&  H 15    &  3711.97& H15    &   2p+ 2P* & 15d+ 2D &  8&  *\\
 3721.85&     1.78&      3.22&  H 14    &  3721.94& H14    &   2p+ 2P* & 14d+ 2D &  8&  *\\
       &         *&         *&  [S III] &  3721.63& F2     &   3p2 3P  &  3p2 1S &  3&  1\\
 3726.00&     6.73&      12.16&  [O II]  &  3726.03& F1     &   2p3 4S* &  2p3 2D*&  4&  4\\
 3728.75&     5.59&      10.08&  [O II]  &  3728.82& F1     &   2p3 4S* &  2p3 2D*&  4&  6\\
 3734.18&     1.77&      3.18&  H 13    &  3734.37& H13    &   2p+ 2P* & 13d+ 2D &  8&  *\\
 3750.04&     2.03&      3.64&  H 12    &  3750.15& H12    &   2p+ 2P* & 12d+ 2D &  8&  *\\
 3754.65&     0.74&      1.32&  O III   &  3754.69& V2     &   3s  3P* &  3p  3D &  3&  5\\
 3759.64&     0.98&      1.74&  O III   &  3759.87& V2     &   3s  3P* &  3p  3D &  5&  7\\
 3770.49&     2.38&      4.22&  H 11    &  3770.63& H11    &   2p+ 2P* & 11d+ 2D &  8&  *\\
 3773.90&     0.33&      0.59&  O III   &  3774.02& V2     &   3s  3P* &  3p  3D &  3&  3\\
 3777.82&     0.16:&     0.28&  Ne II   &  3777.14& V1     &   3s  4P  &  3p  4P*&  2&  4\\
 3791.08&     0.36&      0.63&  O III   &  3791.27& V2     &   3s  3P* &  3p  3D &  5&  5\\
 3797.84&     3.25&      5.70&  H 10    &  3797.90& H10    &   2p+ 2P* & 10d+ 2D &  8&  *\\
 3812.97&     0.35&      0.62&  He II   &  3813.50&  4.19  &   4f+ 2F* & 19g+ 2G & 32&  *\\
 3819.55&     0.63&      1.10&  He I    &  3819.62& V22    &   2p  3P* &  6d  3D &  9& 15\\
 3835.36&     4.77&      8.23&  H 9     &  3835.39& H9     &   2p+ 2P* &  9d+ 2D &  8&  *\\
 3858.14&     0.24:&     0.42&  He II   &  3858.07&  4.17  &   4f+ 2F* & 17g+ 2G & 32&  *\\
 3868.76&     59.20&      100.6&  [Ne III]&  3868.75& F1     &   2p4 3P  &  2p4 1D &  5&  5\\
 3881.59&     0.22&      0.37&  O II    &  3882.19& V12    &   3p  4D* &  3d  4D &  8&  8\\
       &         *&         *&  O II    &  3882.45& V11    &   3p  4D* &  3d  4P &  4&  4\\
       &         *&         *&  O II    &  3883.13& V12    &   3p  4D* &  3d  4D &  8&  6\\
 3888.88&     11.20&      18.85&  H 8     &  3889.05& H8     &   2p+ 2P* &  8d+ 2D &  8&  *\\
       &         *&         *&  He I    &  3888.65& V2     &   2s  3S  &  3p  3P*&  3&  9\\
 3923.65&     0.15&      0.25&  He II   &  3923.48&  4.15  &   4f+ 2F* & 15g+ 2G & 32&  *\\
 3967.58&     17.30&      28.12&  [Ne III]&  3967.46& F1     &   2p4 3P  &  2p4 1D &  3&  5\\
 3970.26&     10.60&      17.19&  H 7     &  3970.07& H7     &   2p+ 2P* &  7d+ 2D &  8& 98\\
 4026.37&     1.51&      2.38&  He I    &  4026.21& V18    &   2p  3P* &  5d  3D &  9& 15\\
 4036.23&     0.12:&     0.19&  N II    &  4035.08& V39a   &   3d  3F* & 4f 2[4] &  5&  7\\
 4070.16&     0.96&      1.49&  [S II]  &  4068.60& F1     &   2p3 4S* &  2p3 2P*&  4&  4\\
 4076.22&     0.21&      0.32&  [S II]  &  4076.35& F1     &   2p3 4S* &  2p3 2P*&  2&  4\\
 4097.95&     1.15&      1.75&  N III   &  4097.33& V1     &   3s  2S  &  3p  2P*&  2&  4\\
 4102.11&     19.50&      29.65&  H 6     &  4101.74& H6     &   2p+ 2P* &  6d+ 2D &  8& 72\\
 4120.53&     0.40&      0.59&  He I    &  4120.84& V16    &   2p  3P* &  5s  3S &  9&  3\\
 4144.12&     0.20&      0.29&  He I    &  4143.76& V53    &   2p  1P* &  6d  1D &  3&  5\\
 4187.46&     0.19&      0.28&  C III   &  4186.90& V18    &   4f  1F* &  5g  1G &  7&  9\\
 4200.23&     0.41&      0.60&  He II   &  4199.83&  4.11  &   4f+ 2F* & 11g+ 2G & 32&  *\\
       &         *&         *&  N III   &  4200.10& V6     &   3s' 2P* &  3p' 2D &  4&  6\\
 4267.56&     0.764&     1.064&  C II   &  4267.15& V6     &   3d  2D  &  4f  2F*& 10& 14\\
 4285.32&     0.13&      0.19&  O II    &  4285.69& V78b   &   3d  2F  &  4f  F3*&  6&  8\\
 4306.65&     0.11&      0.15&  O II    &  4307.23& V53b   &   3d  4P  &  4f  D2*&  2&  4\\
 4340.79&     35.90&      48.09&  H 5     &  4340.47& H5     &   2p+ 2P* &  5d+ 2D &  8& 50\\
 4357.68&     0.29&      0.45&  O II    &  4357.25& V63a   &   3d  4D  &  4f  D3*&  6&  8\\
       &         *&         *&  O II    &  4357.25& V63a   &   3d  4D  &  4f  D3*&  6&  6\\
 4363.53&     7.86&      10.39&  [O III] &  4363.21& F2     &   2p2 1D  &  2p2 1S &  5&  1\\
 4367.55&     0.30&      0.40&  O II    &  4366.89& V2     &   3s  4P  &  3p  4P*&  6&  4\\
 4388.28&     0.45&      0.58&  He I    &  4387.93& V51    &   2p  1P* &  5d  1D &  3&  5\\
 4397.42&     0.08&      0.11&  Ne II   &  4397.99& V57b   &   3d  4F  & 4f 1[4]*&  6&  8\\
 4409.66&     0.09:&     0.11&  Ne II   &  4409.30& V55e   &   3d  4F  & 4f 2[5]*&  8& 10\\
 4413.43&     0.12&      0.16&  Ne II   &  4413.22& V65    &   3d  4P  & 4f 0[3]*&  6&  8\\
       &         *&         *&  Ne II   &  4413.11& V57c   &   3d  4F  & 4f 1[3]*&  4&  6\\
       &         *&         *&  Ne II   &  4413.11& V65    &   3d  4P  & 4f 0[3]*&  6&  6\\
 \noalign{\smallskip}
 \hline
 \end{tabular}
 \end{table*}
 \setcounter{table}{0}
 \begin{table*}
 \tabcolsep 10pt
 \centering
 \caption{{\it --continued}}
 \begin{tabular}{lccccccccc}
 \hline
 \hline
 \noalign{\smallskip}
 $\lambda_{\rm obs}$&$F(\lambda)$&                      $I(\lambda)$&Ion&$\lambda_{\rm lab}$                           &Mult&Lower Term&Upper Term&$g_1$&$g_2$\\
 (1) & (2) & (3) & (4) & (5) & (6)                        & (7) & (8)& (9)& (10) \\
 \noalign{\smallskip}
 \hline
 \noalign{\smallskip}
 4430.71&     0.11:&     0.14&  Ne II   &  4430.94& V61a   &   3d  2D  & 4f 2[4]*&  6&  8\\
 4457.24&     0.10:&     0.12&  Ne II   &  4457.05& V61d   &   3d  2D  & 4f 2[2]*&  4&  6\\
       &         *&         *&  Ne II   &  4457.24& V61d   &   3d  2D  & 4f 2[2]*&  4&  4\\
 4471.74&     3.02&     3.76&  He I    &  4471.50& V14    &   2p  3P* &  4d  3D &  9& 15\\
 4477.54&     0.09:&    0.11&  O II    &  4477.90& V88    &   3d  2P  &  4f  G3*&  4&  6\\
 4480.86&     0.05:&    0.06&  Mg II   &  4481.21& V4     &   3d  2D  &  4f  2F*& 10& 14\\
 4487.70&     0.13&     0.16&  O II    &  4487.72& V104   &   3d' 2P  &  4f' D2*&  2&  4\\
       &         *&         *&  O II    &  4488.20& V104   &   3d' 2P  &  4f' D2*&  4&  6\\
 4510.48&     0.15&     0.18&  N III   &  4510.91& V3     &   3s' 4P* &  3p' 4D &  4&  6\\
       &         *&         *&  N III   &  4510.91& V3     &   3s' 4P* &  3p' 4D &  2&  4\\
 4541.75&     1.04&     1.24&  He II   &  4541.59&  4.9   &   4f+ 2F* &  9g+ 2G & 32&  *\\
 4563.02&     0.16&     0.19&  Mg I]   &  4562.60&        &   3s2 1S  & 3s3p 3P*&  1&  5\\
 4571.23&     0.19&     0.23&  Mg I]   &  4571.10&        &   3s2 1S  & 3s3p 3P*&  1&  3\\
 4610.34&     0.11:&    0.13&  O II    &  4610.20& V92c   &   3d  2D  &  4f  F2*&  4&  6\\
 4613.89&     0.11&     0.13&  N II    &  4613.87& V5     &   3s  3P* &  3p  3P &  3&  3\\
       &         *&         *&  O II    &  4613.14& V92b   &   3d  2D  &  4f  F3*&  6&  6\\
       &         *&         *&  O II    &  4613.68& V92b   &   3d  2D  &  4f  F3*&  6&  8\\
 4621.17&     0.087&     0.10&  N II    &  4621.39& V5     &   3s  3P* &  3p  3P &  3&  1\\
 4634.66&     0.51&     0.58&  N III   &  4634.14& V2     &   3p  2P* &  3d  2D &  2&  4\\
 4640.97&     1.18&     1.34&  N III   &  4640.64& V2     &   3p  2P* &  3d  2D &  4&  6\\
 4647.17&     0.49&     0.56&  C III   &  4647.42& V1     &   3s  3S  &  3p  3P*&  3&  5\\
 4650.76&     0.80&     0.91&  C III   &  4651.47& V1     &   3s  3S  &  3p  3P*&  3&  1\\
 4657.12&     0.20&     0.22&  [Fe III]&  4658.10& F3     &   3d6 5D  &  3d6 3F2&  9&  9\\
        &        *&        *&  C IV    &  4658.30&        &        5  &       6&    &   \\
 4661.37&     0.29&     0.45&  O II    &  4661.63& V1     &   3s  4P  &  3p  4D*&  4&  4\\
 4685.99&     36.50&     40.30&  He II   &  4685.68&  3.4   &   3d+ 2D  &  4f+ 2F*& 18& 32\\
 4697.10&     0.13&     0.14&  O II    &  4696.35& V1     &   3s  4P  &  3p  4D*&  6&  4\\
 4711.72&     2.44&     2.66&  [Ar IV] &  4711.37& F1     &   3p3 4S* &  3p3 2D*&  4&  6\\
 4740.52&     1.87&     2.00&  [Ar IV] &  4740.17& F1     &   3p3 4S* &  3p3 2D*&  4&  4\\
 4861.84&     100.0&     100.0&  H 4     &  4861.33& H4     &   2p+ 2P* &  4d+ 2D &  8& 32\\
 4881.24&     0.061&    0.060 & [Fe III] & 4881.11 & F2   & 3d6 5D  & 3d6 3H    & 9 & 9 \\
 4922.45&     1.11&     1.07&  He I    &  4921.93& V48    &   2p  1P* &  4d  1D &  3&  5\\
 4932.05&     0.37&     0.35&  [O III] &  4931.80& F1     &   2p2 3P  &  2p2 1D &  1&  5\\
 4959.47&     416.0&     393.6&  [O III] &  4958.91& F1     &   2p2 3P  &  2p2 1D &  3&  5\\
 5007.30&     1250.&     1151.&  [O III] &  5006.84& F1     &   2p2 3P  &  2p2 1D &  5&  5\\
 5199.15&     0.930&     0.77&  [N I]   &  5199.84& F1     &   2p3 4S* &  2p3 2D*&  4&  4\\
       &         *&         *&  [N I]   &  5200.26& F1     &   2p3 4S* &  2p3 2D*&  4&  6\\
 5412.04&     5.01&     3.68&  He II   &  5411.52&  4.7   &   4f+ 2F* &  7g+ 2G & 32& 98\\
 5453.01&     0.21:&    0.15&  S II    &  5453.83& V6     &   4s  4P  &  4p  4D*&  6&  8\\
 5518.04&     1.09&     0.76&  [Cl III]&  5517.66& F1     &   2p3 4S* &  2p3 2D*&  4&  6\\
 5538.04&     1.04&     0.72&  [Cl III]&  5537.60& F1     &   2p3 4S* &  2p3 2D*&  4&  4\\
 5666.98&     0.19&     0.12&  N II    &  5666.63& V3     &   3s  3P* &  3p  3D &  3&  5\\
 5674.72&     0.23&     0.15&  N II    &  5676.02& V3     &   3s  3P* &  3p  3D &  1&  3\\
 5755.87&     0.39&     0.25&  [N II]  &  5754.60& F3     &   2p2 1D  &  2p2 1S &  5&  1\\
 5802.58&     0.71&     0.45&  C IV    &  5801.51& V1     &   3s  2S  &  3p  2P*&  2&  4\\
 5811.60&     0.72&     0.45&  C IV    &  5812.14& V1     &   3s  2S  &  3p  2P*&  2&  2\\
 5876.14&     18.70&     11.40&  He I    &  5875.66& V11    &   2p  3P* &  3d  3D &  9& 15\\
 6037.34&     0.53&     0.31&  He II   &  6036.70&  5.21  &   5g+ 2G  & 21h+ 2H*& 50&  *\\
 6074.27&     0.37&     0.21&  He II   &  6074.10&  5.20  &   5g+ 2G  & 20h+ 2H*& 50&  *\\
 6102.35&     0.76&     0.43&  [K IV]  &  6101.83& F1     &   3p4 3P  &  3d4 1D &  5&  5\\
 6118.58&     0.52&     0.29&  He II   &  6118.20&  5.19  &   5g+ 2G  & 19h+ 2H*& 50&  *\\
 6235.04&     0.40&     0.21&  He II   &  6233.80&  5.17  &   5g+ 2G  & 17h+ 2H*& 50&  *\\
 6300.45&     0.27&     0.14&  [O I]   &  6300.34& F1     &   2p4 3P  &  2p4 1D &  5&  5\\
 6312.51&     2.85&     1.48&  [S III] &  6312.10& F3     &   2p2 1D  &  2p2 1S &  5&  1\\
       &         *&         *&  He II   &  6310.80&  5.16  &   5g+ 2G  & 16h+ 2H*& 50&  *\\
 6393.54&     0.13&     0.07&  [Mn V]  &  6393.60&        &   3d3 4F  &  3d3 4P & 10&  6\\
 6407.79&     0.48&     0.24&  He II   &  6406.30&  5.15  &   5g+ 2G  & 15h+ 2H*& 50&  *\\
 6527.78&     0.60&     0.29&  He II   &  6527.11&  5.14  &   5g+ 2G  & 14h+ 2H*& 50&  *\\
 6548.69&     6.57&     3.16&  [N II]  &  6548.10& F1     &   2p2 3P  &  2p2 1D &  3&  5\\
 6563.26&     665.0&     318.3&  H 3     &  6562.77& H3     &   2p+ 2P* &  3d+ 2D &  8& 18\\
 6583.95&     20.90&     9.93&  [N II]  &  6583.50& F1     &   2p2 3P  &  2p2 1D &  5&  5\\
 \noalign{\smallskip}
 \hline
 \end{tabular}
 \end{table*}
 \setcounter{table}{0}
 \begin{table*}
 \tabcolsep 10pt
 \centering
 \caption{{\it --continued}}
 \begin{tabular}{lccccccccc}
 \hline
 \hline
 \noalign{\smallskip}
 $\lambda_{\rm obs}$&$F(\lambda)$&                      $I(\lambda)$&Ion&$\lambda_{\rm lab}$                           &Mult&Lower Term&Upper Term&$g_1$&$g_2$\\
 (1) & (2) & (3) & (4) & (5) & (6)                        & (7) & (8)& (9)& (10) \\
 \noalign{\smallskip}
 \hline
 \noalign{\smallskip}
 6678.73&     6.97&     3.21&  He I    &  6678.16& V46    &   2p  1P* &  3d  1D &  3&  5\\
 6717.04&     3.90&     1.77&  [S II]  &  6716.44& F2     &   2p3 4S* &  2p3 2D*&  4&  6\\
 6731.42&     4.31&     1.95&  [S II]  &  6730.82& F2     &   2p3 4S* &  2p3 2D*&  4&  4\\
 6795.87&     0.28&     0.12&  [K IV]  &  6795.00& F1     &   3p4 3P  &  3p4 1D &  3&  5\\
 6891.36&     0.84&     0.36&  He I    &  6890.88& 5.12   &   5g+ 2G  & 12h+ 2H & 50&  *\\
 7065.82&     6.12&     2.51&  He I    &  7065.25& V10    &   2p  3P* &  3s  3S &  9&  3\\
 7136.34&     35.70&     14.34&  [Ar III]&  7135.80& F1     &   3p4 3P  &  3p4 1D &  5&  5\\
 7159.16&     0.11&     0.045&  He I    &  7160.56&        &   3s  3S  & 10p  3P*&  3&  9\\
 7178.31&     1.11&     0.44&  He II   &  7177.50&  5.11  &   5g+ 2G  & 11h+ 2H*& 50&  *\\
 7231.51&     0.34&     0.13&  C II    &  7231.32& V3     &   3p  2P* &  3d  2D &  2&  4\\
 7237.20&     0.92&     0.36&  C II    &  7236.42& V3     &   3p  2P* &  3d  2D &  4&  6\\
       &         *&         *&  C II    &  7237.17& V3     &   3p  2P* &  3d  2D &  4&  4\\
       &         *&         *&  [Ar IV] &  7237.26& F2     &   3p3 2D* &  3p3 2P*&  6&  4\\
 7262.88&     0.21&     0.08&  [Ar IV] &  7262.76& F2     &   3p3 2D* &  3p3 2P*&  4&  2\\
 7282.00&     1.42&     0.55&  He I    &  7281.35& V45    &   2p  1P* &  3s  1S &  3&  1\\
 7320.36&     1.91&     0.73&  [O II]  &  7318.92& F2     &   2p3 2D* &  2p3 2P*&  6&  2\\
       &         *&         *&  [O II]  &  7319.99& F2     &   2p3 2D* &  2p3 2P*&  6&  4\\
 7330.69&     1.55&     0.59&  [O II]  &  7329.67& F2     &   2p3 2D* &  2p3 2P*&  4&  2\\
       &         *&         *&  [O II]  &  7330.73& F2     &   2p3 2D* &  2p3 2P*&  4&  4\\
       &         *&         *&  [Ar IV] &  7331.40& F2     &   3p3 2D* &  3p3 2P*&  6&  2\\
 7531.01&     0.72&     0.26&  [Cl IV] &  7530.54&        &   3p2 3P  &  3p2 1D &  3&  5\\
 7592.85&     1.28&     0.45&  He II   &  7592.74& 5.10   &   5g+ 2G  &  10h+ 2H&  5&  *\\
 7618.89&     0.78&     0.27&  N V     &  7618.46& V13    &   7g+ 2G  &   8h+ 2H&  *&  *\\
 7751.64&     9.84&     3.52&  [Ar III]&  7751.06&        &   3p4 3P  &  3p4 1D &  3&  5\\
 8046.37&     1.70&     0.54&  [Cl IV] &  8045.63&        &   3p2 3P  &  3p2 1D &  5&  5\\
 \noalign{\smallskip}
 \hline
 \end{tabular}
 \end{table*}

\subsection{The Central Star Spectrum}
\label{sub:cs1501}
\begin{figure*}
\vspace{0.4cm}
 \centering \epsfig{file=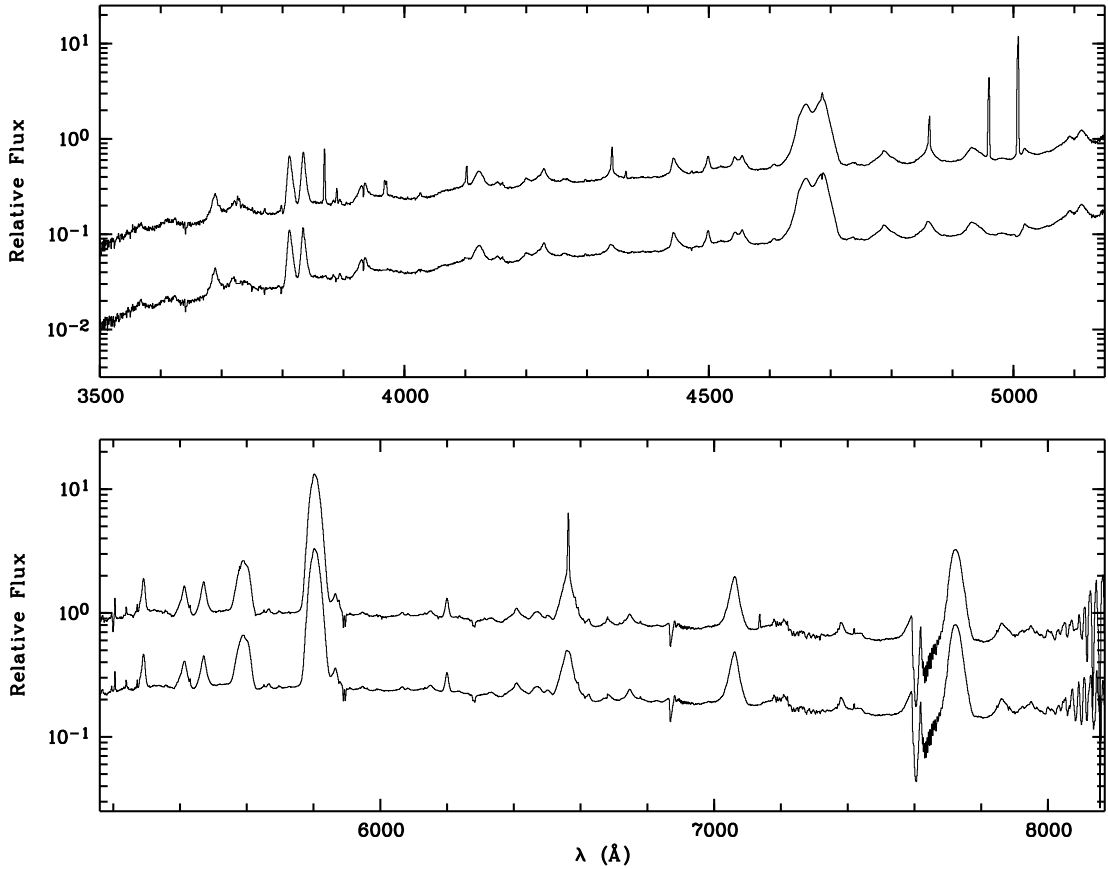,
height=11.7cm, bbllx=39, bblly=227, bburx=546, bbury=629, clip=, angle=0}
\caption{Observed optical spectrum of the central star. Each panel shows the spectrum detected in the centre of the slit 
(upper line) and the same spectrum after subtraction of the nebular spectrum (lower line). Due to atmospheric dispersion effects on our narrow-slit spectrum, the slope of the stellar energy distribution is not reliable.}
\label{fig:spectrum_cs}
\end{figure*}

 \begin{table*}
 \tabcolsep 10pt
 \centering
 \caption{\label{tab:cslines} Identifications and measured equivalent widths (EW) of emission features in the 
spectrum of the central star of NGC~1501. The EWs were derived after subtraction of the nebular spectrum.}
 \vspace{0.3cm}
 \begin{tabular}{llllll}
 \hline
 \hline
 \noalign{\smallskip}
 $\lambda_{\rm obs}$& EW & Ion & $\lambda_{\rm lab}$ &Lower Term&Upper Term\\
 ({\AA})    & ({\AA})  &       &       ({\AA})       &          &          \\
 \noalign{\smallskip}
 \hline
 \noalign{\smallskip}
 3566       & 1.5     & C~{\sc iv} &  3566.8         & 7        & 15 \\
 3620       & 3.8     & C~{\sc iv} &  3607.3         & 6p       & 9d \\
            & *       & O~{\sc vi} &  3615.6         & 6p       & 7s \\      
 3689       & 7.9     & C~{\sc iv} &  3689.7         & 6        & 9 \\
 3723       & 11.2    & C~{\sc iv} &  3720.1         & 7        & 14 \\
            & *       & O~{\sc v}  &  3698-3725      & 3p$^\prime~^3$D & 3d$^\prime~^3$D$^{\rm o}$ \\
 3811       & 24.5    & O~{\sc vi} &  3811.4         & 3s~$^2$S & 3p~$^2$P$^{\circ}$\\
 3834       & 21.2    & O~{\sc vi} &  3834.4         & 3s~$^2$S & 3p~$^2$P$^{\circ}$\\
 3933       & 11.0    & C~{\sc iv} &  3929.3         & 7        & 13 \\
            & *       & C~{\sc iv} &  3934.4         & 5s~$^2$S & 6p~$^2$P$^{\circ}$\\
            & *       & O~{\sc vi} &  3936.6         &  9       & 13 \\
 3976       & 5.0     & ?          &                 &          &    \\
 4026       & 0.70    & He~{\sc ii}&  4025.6         &  4       & 13 \\
 4121       & 6.2     & O~{\sc v}  &  4119-4124      & 3s$^\prime~^3$P$^{\circ}$& 3p$^\prime~^3$D\\
 4153       & 1.6     & O~{\sc v}  &  4153.3         & 3s$^\prime~^3$P$^{\circ}$& 3p$^\prime~^3$D\\
 4161       & *       & C~{\sc iv} &  4160.2         & 6p       & 12d \\
 4199       & 1.26    & He~{\sc ii}&  4199.8         & 4        & 11  \\
 4221       & 5.0     & C~{\sc iv} &  4219.0         & 6s~$^2$S & 8p~$^2$P$^{\circ}$\\
 4229       & *       & C~{\sc iv} &  4229.1         & 7        & 12  \\
 4265       & 0.94    & ?          &                 &          &     \\
 4298       & 0.17    & ?          &                 &          &     \\
 4340       & 4.2     & He~{\sc ii}&  4338.7         & 4        & 10  \\
 4412       & 0.23    & ?          &                 &          &     \\ 
 4442       & 6.8     & C~{\sc iv} &  4441.1         & 5p       & 6d  \\
 4479       & 0.14    & O~{\sc v}  &  4479.5         & 6g~$^3$G & 7h~$^3$H \\
 4499       & 2.8     & O~{\sc vi} &  4499.0         & 8        & 10  \\
 4520       & 1.5     & O~{\sc vi} &  4521.2         & 7        & 9   \\
 4541       & 3.0     & He~{\sc ii}&  4541.6         & 4        & 9   \\
 4554       & 4.0     & C~{\sc iv} &  4554.3         & 6p       & 8d  \\
            & *       & O~{\sc v}  &  4554.5         & 3p$^\prime~^1$P & 3d$^\prime~^1$D  \\
 4607       & 0.65    & C~{\sc iv} &  4604.6         & 7p       & 11d  \\
 4659       & 85      & C~{\sc iv} &  4658.7         &   5      &   6  \\  
 4686       & 107     & He~{\sc ii}&  4685.7         &   3      &   4  \\
  *         & *       & C~{\sc iv}&   4685.4         &   6      &   8  \\
  *         & *       & C~{\sc iv}&   4688.9         &   7      &   11 \\
 4736       & 0.60    & C~{\sc iv}&   4736.0         &   6d     &   8p \\    
 4780       & 9.8     & O~{\sc vi}&   4773.3         &   7s     &   8p \\ 
            & *       & C~{\sc iv}&   4785.9         &   5d     &   6f \\
            & *       & C~{\sc iv}&   4789.3         &   6p     &   8s \\
 4861       & 9.1     & He~{\sc ii}&  4859.3         &   4      &   8  \\
            & *       & C~{\sc iv} &  4860           &   8      &   16 \\
 4931       & 9.4     & O~{\sc v}  &  4930.3         &   6h     &   7i          \\
 4981       & 0.66    & O~{\sc v}  &  4977-4984      & 6f~$^3$F$^{\rm o}$ & 7g~$^3$G \\  
 5018       & 3.1     & C~{\sc iv} &  5017.8         & 5p~$^2$P$^{\circ}$ & 6s~$^2$S\\
 5092       & 13.0    & C~{\sc iv}&   5092.7         &   8      &   15 \\
            & *       & O~{\sc vi}&   5083.9         &   7p     &   8d \\
 5114       & *       & O~{\sc v} &   5114.1         & 3s~$^1$S  & 3p~$^1$P$^{\circ}$\\
 5168       & 0.81    & O~{\sc vi}&   5166.7         &  10      &   14 \\ 
 5291       & 10.4    & O~{\sc vi}&   5290.6         &   7      &    8 \\
 5325-5358  & 1.9     & C~{\sc iv}&   5335.2         &   8p     &  14d \\
            & *       & C~{\sc iv}&   5356.9         &   7p     &  10d \\ 
 5413       & 15.1    & He~{\sc ii}&  5411.5         &   4      &    7 \\
            & *       & C~{\sc iv} &  5411.0         &   8      &   14 \\           
 5471       & 15.8    & C~{\sc iv} &  5470.7         &   7      &   10 \\     
 5518-5538  & 0.95    & C~{\sc iv} &  5518.8         &   7p     &   10s \\
            & *       & C~{\sc iv} &  5530.5         &   8s     &   13p \\    
 5591       & 65.5    & O~{\sc v}  &  5572-5604      & 3p~$^3$P$^{\circ}$& 3d~$^3$D \\ 
 5651       & 0.58    &            &                 &          &      \\
 5667       & 1.05    & O~{\sc v}  &  5668.8         &   9      &   13 \\
            & *       & O~{\sc vii}&  5669.3         &   8      &    9 \\
 \noalign{\smallskip}
 \hline
 \end{tabular}
 \end{table*}
 \setcounter{table}{1}
 \begin{table*}
 \tabcolsep 10pt
 \centering
 \caption{{\it --continued}}
 \begin{tabular}{llllll}
 \hline
 \hline
 \noalign{\smallskip}
 $\lambda_{\rm obs}$& EW & Ion & $\lambda_{\rm lab}$ &Lower Term&Upper Term\\
 ({\AA})    & ({\AA})  &       &       ({\AA})       &          &          \\
 \noalign{\smallskip}
 \hline
 \noalign{\smallskip}
 5696       & 0.40    & C~{\sc iii}&  5695.9         & 3p~$^1$P$^{\circ}$& 3d~$^1$D \\
 5806       & 448     & C~{\sc iv} &  5801.3,5812.0  & 3s~$^2$S  & 3p~$^2$P$^{\circ}$\\
 5865       & 7.0     & C~{\sc iv} &  5865.4         &   8      &   13  \\
 5947       & 1.46    & C~{\sc iv} &  5946.9         &   9      &   19  \\
 6066       & 0.88    & O~{\sc viii}& 6068.2         &   9      &   10  \\
            & *       & O~{\sc viii}& 6064.2         &  11      &   13  \\
 6084       & 0.34    & O~{\sc vii} & 6085.1         &  10      &   12  \\
 6120       & 0.28    & He~{\sc ii} & 6118.3         &   5      &   19  \\
 6136       & 0.36    & O~{\sc v}   & 6131           & 4s$^\prime~^3$P$^{\rm o}$ & 4p$^\prime~^3$P \\
 6150-6170  & 1.9     & C~{\sc iv}  & 6149.9         &   8s     &   12p \\
            & *       & He~{\sc ii} & 6170.7         &   5      &   18  \\
 6200       & 4.8     & O~{\sc vi}  & 6200.8         &   9      &   11  \\
  *         & *       & O~{\sc vi}  & 6198.3         &  10      &   13  \\
 6235       & 0.46    & He~{\sc ii} & 6233.8         &   5      &   17  \\
 6332       & 0.85    & O~{\sc v}   & 6330.1         & 3p$^\prime~^1$D & 3d$^\prime~^1$F$^{\rm o}$ \\ 
 6408       & 8.4     & C~{\sc iv}  & 6404.4         &   7s     &    9p \\
            & *       & C~{\sc iv}  & 6408.7         &   9      &   17  \\ 
            & *       & He~{\sc ii} & 6406.4         &   5      &   15  \\
 6461       & 4.3     & O~{\sc v}   & 6460,66        & 3p~$^3$D     & 3d~$^3$F$^{\circ}$\\
 6502       & 1.35    & O~{\sc v}   & 6500.2         & 3p~$^3$D     & 3d~$^3$F$^{\circ}$\\
 6563       & 51.4    & He~{\sc ii} & 6560.1         &   4      &    6  \\
            & *       & C~{\sc iv}  & 6559.5         &   8      &   12  \\
 6624       & 1.2     &             &                &          &       \\
 6682       & 2.4     & He~{\sc ii} & 6683.2         &   5      &   13  \\
 6747       & 4.3     & O~{\sc v}   & 6747.0         &   9      &   12  \\
  *         & *       & C~{\sc iv}  & 6747.5         &   9      &   16  \\
 7062       & 53.0    & C~{\sc iv}  & 7062.4         &   7      &    9  \\
 7382       & 4.3     & C~{\sc iv}  & 7380.4         &   6p     &    7d \\
 7434       & 2.2     & O~{\sc v}   & 7422-7443      & 4s$^\prime~^3$S & 4p$^\prime~^3$P$^{\rm o}$ \\  
 7723       & 187     & C~{\sc iv}  & 7726.4         &   6      &    7  \\
  *         & *       & C~{\sc iv}  & 7735.9         &   8      &   11  \\
  *         & *       & O~{\sc vi}  & 7717.1         &   8      &    9  \\
 7862       & 12.0    & C~{\sc iv}  & 7860.7         &   9      &   14  \\
 7924       & 1.5     & O~{\sc vii} & 7926.4         &   9      &   10  \\
 7949       & 7.1     & C~{\sc iv}  & 7947.2         &   6d     &    7p \\
 \hline
 \end{tabular}
 \end{table*}

The upper and lower panels of Figure~\ref{fig:spectrum_cs} show the blue and red spectra measured at the central star position. In each panel the upper plot shows the unsubtracted spectrum and the lower shows the spectrum after subtraction of the nebular spectrum. The central star spectrum was integrated over 5.8\,arsec (30 pixels) along the slit, and the nebular spectrum subtracted from it was taken an equal area 10\,arcsec (50 pixels) away from the central position. The Figure shows that some emission lines have both a stellar and a nebular component, but most are only produced by one or the other. As expected, broadening of the stellar lines is evident, while the width of the detected nebular lines is much narrower. 

Our 2.5~hour exposure on the 14th magnitude central star of NGC~1501 with the 4.2-m WHT's ISIS spectrograph, led to a very high signal to noise spectrum. The stellar emission lines are relatively narrow for a star of its class, and so many lines could be identified. Emission lines from four ionization stages of oxygen (O~{\sc v} -- O~{\sc viii}), together with numerous C~{\sc iv} and He~{\sc ii} lines, were found to be present. We detected a very weak C~{\sc iii} 5696~\AA\ emission feature, whose EW of 0.40~\AA\ was 1120 times weaker than that of the C~{\sc iv} 5801,12~\AA\ doublet.  Identifications and measured equivalent widths (EWs) for the stellar emission lines are listed in Table~2. Our EWs show quite good agreement with those measured for a subset of the lines by Stanghellini, Kaler \& Shaw (1994), though we note that our ionic attributions for the lines at 3689, 3933, 4780, 4931, 5291 and 5471~\AA\ differ from those of Stanghellini et al.  Using the WO star classification criteria listed in Table~3 of Crowther et al. 1998, the EWs listed in our Table~2 yield a spectral type of WO4, in agreement with the subtype assigned by Crowther et al. to this central star.

The airmass at the time of the observations was quite large (1.44-2.13), so that the effects of differential atmospheric dispersion on the narrow (0.75~arcsec) slit needed for the nebular ORL analysis led to significant losses towards the blue end of the central star spectrum. For this reason, neither the spectral slope nor the absolute continuum fluxes of the central star of NGC 1501 can be derived from the optical observations presented in this paper.

Despite the g-mode pulsations detected in the central star of NGC~1501 \citep{bond96} the emission-line optical spectrum of this star (see Figure~\ref{fig:spectrum_cs}) points towards the presence of a thick wind - no stellar absorption lines were detected at all. \citet{koesteke97} derived a mass-loss-rate of 5.2$\cdot$10$^{-7}$~M$_{\odot}$yr$^{-1}$ and a terminal  stellar wind velocity of 1800~km\,s$^{-1}$, while \citet{feibelman98} estimated the terminal stellar wind velocity to be in the 3260 to 3460\,km\,s$^{-1}$ range from its IUE low-resolution spectrum. Following the method of \citet{kingsburgh95}, recently reviewed by \citet{drew04}, we estimate the terminal stellar wind velocity by measuring the half width at zero intensity (HWZI) of the C~{\sc iv}~$\lambda\lambda$5801,12 doublet. \citet{kingsburgh95} showed that in the case of Sand~1, a SMC W03 star, the velocity corresponding to the HWZI of C~{\sc iv}~$\lambda\lambda$5801,12 doublet, corrected for the 10.65~{\AA} doublet splitting, was consistent with the terminal wind velocity measured from the black absorption edge of the C~{\sc iv}~$\lambda\lambda$1548,51 resonance doublet, measured in a high resolution $IUE$ spectrum. The validity of the method was later confirmed for several other objects. We measured a full width at zero intensity (FWZI) of 103~{\AA} for the C~{\sc iv}~$\lambda\lambda$5801,12 doublet, which, following the \citet{kingsburgh95} approach discussed above, yields a corrected HWZI of 46.2~{\AA} and a terminal stellar wind velocity of 2390~km\,s$^{-1}$ for the central star of NGC~1501. 

\subsection{Reddening}
\label{sub:reddening}
The logarithmic extinction at H${\beta}$, $c({\rm H}{\beta})= {\rm log}[I({\rm H}\beta)/F({\rm H}{\beta})]$, was derived from the Balmer line ratios, H${\alpha}/{\rm H}{\beta}$,  H${\gamma}/{\rm H}{\beta}$, H${\delta}/{\rm H}{\beta}$ and H${\epsilon}/{\rm H}{\beta}$. The measured ratios, together with the corresponding value of $c({\rm H}{\beta})$ are given in Table~\ref{tab:reddening}. The theoretical line ratios used were those given by \citet{storey95} for $T_{\rm e}$~=~11~kK and $N_{\rm e}$~=~10$^3$~cm$^{-3}$ and we adopted the Galactic reddening law of \citet{howarth83}. Given the uncertainty introduced by the different flux calibrations for the two arms of the spectrograph, we expect large error bars on our measured H$\alpha$/H$\beta$ ratio and on the corresponding value of c(H${\beta}$). We finally adopted the value c(H$\beta$)\,=\,1.0 as this best corrected the 2200~{\AA} interstellar absorption feature. This value, which is used to correct observed fluxes for interstellar extinction throughout this work, is within the range of values reported in the literature for this object -~0.96: \citet{kaler76}; 1.2: \citet{cahn92}; 1.1: \citet{stanghellini94}; 1.11: \citet{ciardullo96}; 1.05: \citet{sabbadin00a}. 

\begin{table}
\caption{Logarithmic extinction coefficient derived from the Balmer decrement}
\label{tab:reddening}
\centering
\begin{tabular}{lccc}
\hline
\hline
\noalign{\smallskip}
 Balmer decrement             &\multicolumn{2}{c}{Ratio} & c(H$\beta$) \\
                              & Theory$^a$ & Observed    &                \\
 \noalign{\smallskip}
 \hline
 \noalign{\smallskip}
 H${\alpha}/{\rm H}{\beta}$ &  2.837     & 6.65        & 1.15           \\
 H${\gamma}/{\rm H}{\beta}$ &  0.470     & 0.359       & 0.92           \\
 H${\delta}/{\rm H}{\beta}$ &  0.260     & 0.195       & 0.69           \\
 H${\eta}/{\rm H}{\beta}$   &  0.160     & 0.106       & 0.86           \\
 Adopted                      &            &             & 1.0            \\
\noalign{\smallskip}
\hline
\end{tabular}
\\
\small{$^a$Theoretical line ratios are from Storey \& Hummer (1995).}
\end{table}

\subsection{A recombination line abundance analysis for the WO4 central
star}
\label{sub:csabundances}

The WO4 central star of NGC~1501 has a pure emission line spectrum and so we have used the recombination line method of \citet[][ KBS95]{kingsburgh95} to derive the relative abundances of helium, carbon and oxygen ions in its wind. For the case of the LMC WO3 star Sanduleak~2, \citet{crowther00} derived He, C and O abundances for its wind via spherical NLTE line-blanketed modelling, which showed good agreement with those derived from the recombination line analysis of KBS95, so we expect our recombination line abundances for the nucleus of NGC~1501 to be reliable.

To overcome the lack of a spectrophotometric calibration for our central star spectrum, we derived dereddened emission line fluxes by normalising the equivalent widths listed in Table~2 to a continuum flux slope corresponding to the model atmospheres discussed in Sections~4.4 and 4.5. The T$_{\rm eff}$=110~kK, log~g=6.0 plane parallel NLTE T\"ubingen model discussed in Section 4.4 has a flux distribution between 4000~\AA\ and 7000~\AA\ which can be aproximated by F$_{\lambda} \propto \lambda^{-n}$, with n=3.93, while the spherically symmetric CMFgen model discussed in Section~4.5 has n=3.85. We adopted a spectral slope of n=3.90 and normalised it to F$_{\lambda}$ = 4.58$\times10^{-14}$ ergs~cm$^{-2}$~s$^{-1}$~\AA $^{-1}$ at 5470~\AA , corresponding to dereddening by c(H$\beta$)~=~1.00 the continuum-level V magnitude of 14.44 measured by Stanghellini et al. (1994). Column 3 of Table~4 lists the resulting dereddened line fluxes used for the abundance analysis, while column 4 lists the adopted line emission coefficients, Q (= h$\nu\alpha_{rec}$), for T$_e$~=~5$\times10^4$~K and N$_e$~=~10$^{11}$~cm$^{-3}$, taken from Table~14 of KBS95. Column 5 lists the ionic emission measures, I/Q, derived from each line, as well as mean I/Q values for each ion when more than one line was available.  The C$^{4+}$ emission measures derived from the four C~{\sc iv} recombination lines used for the abundance analysis show excellent agreement with each other. Their mean I/Q value was used to predict the fluxes of the various C~{\sc iv} lines that blend with the He~{\sc ii} lines used for our analysis. After correcting for these blends, the I/Q's derived from the He~{\sc ii} lines showed excellent inter-agreement.

Summing all ionic species, we find number ratios of C/He~=~0.44, O/He~=~0.107 and (C+O)/He~=~0.55 for the central star of NGC~1501. These ratios are not dissimilar to the number ratios of C/He~=~0.48 and O/He~=~0.054 derived by \citet{barlow82} from a recombination line analysis of the WO1 PN central star Sanduleak~3. Expressed as mass fractions, we obtain He:C:O = 0.36:0.48:0.16 for the central star of NGC~1501, very similar to the values that have been derived from absorption-line analyses of pulsating PG~1159-type objects, stars which have been postulated to be the descendants of PN WC/WO central stars (see Werner, Heber and Hunger, 1991).

 \begin{table*}
 \tabcolsep 10pt
 \centering
 \caption{\label{tab:cslines_ab} Wind abundance analysis for the central star of NGC~1501. Column 1 lists
the recombining ion, column 2 lists the transition and column 3 lists the dereddened line flux obtained by normalising
the equivalent width listed in Table~\ref{tab:cslines} to a stellar continuum that
has a flux of F$_\lambda$ = 4.58$\times^{-14}$ ergs~cm$^{-2}$~s$^{-1}$~\AA $^{-1}$ at 5470~\AA\ and a 
spectral slope corresponding to F$_\lambda \propto \lambda^{-3.90}$ (see text). Column 4 lists
the line recombination coefficient Q for T$_{\rm e}$~=~5$\times10^4$~K and N$_{\rm e}$~=~10$^{11}$~cm$^{-3}$ 
(see Kingsburgh, Barlow \& Storey 1995), while column 5 lists the
derived ionic emission measure, I/Q.  Values appearing in brackets are where He~{\sc ii} line fluxes have been
corrected for the contribution made by blended C~{\sc iv} lines, where use has been made of the mean I/Q
derived for C$^{4+}$.}
 \vspace{0.3cm}
 \begin{tabular}{lllll}
 \hline
 \hline
 \noalign{\smallskip}
 Recombining Ion & Transition & I (ergs~cm$^{-2}$~s$^{-1}$) & Q (ergs~cm$^3$~s$^{-1}$) & I/Q (cm$^{-5}$) \\
 \noalign{\smallskip}
 \hline
 \noalign{\smallskip}
  C$^{4+}$  & 4658 (6$\rightarrow$5)  & 7.29$\times10^{-12}$ & 6.65$\times10^{-25}$ & 1.096$\times10^{13}$ \\
            & 5471 (10$\rightarrow$7) & 7.21$\times10^{-13}$ & 6.06$\times10^{-26}$ & 1.195$\times10^{13}$ \\
  & 5801,12 (3d$^2$P$^{\rm o}\rightarrow$3p$^2$S) & 1.626$\times10^{-11}$ & 1.39$\times10^{-24}$ & 1.169$\times10^{13}$ \\
            & 7062 (9$\rightarrow$7)  & 8.96$\times10^{-13}$ & 8.30$\times10^{-26}$ & 1.080$\times10^{13}$ \\
            &   ${\rm Mean}$ C$^{4+}$   &                      &                      & 1.135$\pm$0.05$\times10^{13}$ \\
 \noalign{\bigskip}
  He$^{2+}$ & 4686 (4$\rightarrow$3)  & 8.96$\times10^{-12}$ &                      &                      \\
            &        & $\rightarrow$6.43$\times10^{-12}$ & 2.38$\times10^{-25}$ & 2.70$\times10^{13}$ \\ 
  & (C$^{4+}$ 4685 (8$\rightarrow$6), 4689 (11$\rightarrow$7))  & (2.53$\times10^{-12}$) & 2.23$\times10^{-25}$
& (1.135$\times10^{13}$) \\  
 \noalign{\smallskip}
    & 5412 (7$\rightarrow$4)  & 7.21$\times10^{-12}$ &                      &                   \\
            &            & $\rightarrow$5.47$\times10^{-12}$ & 2.20$\times10^{-26}$ & 2.49$\times10^{13}$ \\
  & (C$^{4+}$  5411 (14$\rightarrow$8)) & (1.74$\times10^{-13}$) & 1.53$\times10^{-26}$ &
(1.135$\times10^{13}$) \\
 \noalign{\smallskip}
   & 6560 (6$\rightarrow$4)  & 1.16$\times10^{-12}$ & & \\
            &   & $\rightarrow$8.40$\times10^{-13}$ & 3.28$\times10^{-26}$ & 2.56$\times10^{13}$ \\ 
 & (C$^{4+}$ 6560 (12$\rightarrow$8)) & (3.20$\times10^{-13}$) & 2.82$\times10^{-26}$ &
(1.135$\times10^{13}$) \\  
 \noalign{\smallskip}
    & ${\rm Mean}$ He$^{2+}$            &                 &              & 2.58$\pm$0.12$\times10^{13}$ \\
 \noalign{\bigskip}
 O$^{5+}$   & 4930 (7$\rightarrow$6)  & 6.45$\times10^{-13}$   & 8.53$\times10^{-25}$ & 7.56$\times10^{11}$ \\
    & 5591 (3d~$^3$D$\rightarrow$3p~$^3$P$^{\circ}$)&2.75$\times10^{-12}$ & 4.33$\times10^{-24}$ & 6.35$\times10^{11}$ \\
    & ${\rm Mean}$  O$^{5+}$            &                 &              & 6.96$\pm$0.60$\times10^{11}$ \\
 \noalign{\bigskip}
 O$^{6+}$   & 5291 (8$\rightarrow$7)  & 5.42$\times10^{-13}$   & 1.05$\times10^{-24}$ & 5.16$\times10^{11}$ \\ 
    & 6200 (11$\rightarrow$9, 13$\rightarrow$10) & 1.35$\times10^{-13}$ & 3.06$\times10^{-24}$ & 4.41$\times10^{11}$ \\
    & ${\rm Mean}$ O$^{6+}$         &                 &              & 4.77$\pm$0.37$\times10^{11}$ \\    
 \noalign{\bigskip}
 O$^{7+}$   & 6085 (12$\rightarrow$10)  & 1.028$\times10^{-14}$ & 3.02$\times10^{-25}$ & 3.40$\times10^{10}$ \\
 \noalign{\bigskip}
 O$^{8+}$ &  6068 (10$\rightarrow$9), 6064 (13$\rightarrow$11) & 2.69$\times10^{-14}$ & 1.72$\times10^{-24}$ & 1.56$\times10^{10}$ \\

 \hline
 \end{tabular}
 \end{table*}

\subsection{Nebular Empirical Abundance Determination} 
\label{sub:plasma}

A number of CELs, which are useful as nebular $N_{\rm e}$ and  $T_{\rm e}$ diagnostics, as well as for abundance determination, were detected in the spectra of NGC~1501. Using multi-level ($\ge$5) atomic models and solving the equations of statistical equilibrium, temperatures and densities were derived from these various CEL line ratios.  The diagnostic line ratios and corresponding values of $N_{\rm e}$ and $T_{\rm e}$ are given in Table~\ref{tab:diagnostics}.

The [S~{\sc ii}]~($\lambda$6716/$\lambda$6731), [O~{\sc ii}]~($\lambda$3729/$\lambda$3726), [Cl~{\sc iii}]~($\lambda$5537/$\lambda$5517) and  [Ar~{\sc iv}]~($\lambda$4740/$\lambda$4711) diagnostic line ratios yield electron densities which are in good agreement with each other, giving an average density and standard error of (1040$\pm$200)\,cm$^{-3}$. A higher density (2400\,cm$^{-3}$) is derived from the [O~{\sc ii}]~($\lambda$3727/$\lambda$7325) ratio. As a result of the higher critical density of the $\lambda\lambda$7320,30 doublet, the density derived from this diagnostic is weighted toward high density regions of the nebula.

The [O~{\sc iii}] nebular-to-auroral line ratio gives a temperature of 11.1~kK, which is slightly lower than the 12.6~kK given by the [N~{\sc ii}] nebular-to-auroral ratio. Results from the photoionisation modelling in Section~5 show that $T_{\rm e}$([O~{\sc iii}]) is almost equal to  $T_{\rm e}$([O~{\sc ii}]) and  $T_{\rm e}$([N~{\sc ii}]), and therefore the elevated temperatures measured from the [O~{\sc ii}] and [N~{\sc ii}] line ratios could be due to recombination excitation of the relevant CELs \citep{rubin86}. 

We calculated the contribution of recombination excitation to the [O~{\sc ii}] $\lambda$7320,7330 and $\lambda$3727,3729 lines using Equation~2 from Liu et al (2000), and the relation I$_{\rm R}$($\lambda$3727,3729)=7.7\,I$_{\rm R}$($\lambda$7320,7330) at 10~kK.  Using the O$^{2+}$/H$^{+}$ abundance derived from CELs (see Section 3.6), the corrected 7325{\AA}/3727{\AA} line ratio is 0.0514, giving a revised temperature of 13.7~kK, compared to the value of 16\,200\,K that is derived when no corrections are made. 

Using the formalism of \citet{liu01b}, we derived $T_{\rm e}$(BJ)~=~9.4~kK from the ratio of the nebular continuum Balmer discontinuity at 3646~{\AA} to H11~$\lambda$3770. We note that the signal-to-noise ratio in this region of the spectrum is rather poor (see Figure~1) and so the error on $T_{\rm e}(BJ)$ is quite large.  Also, scattered starlight affects the measured magnitude of the Balmer jump.  However, by comparison of the flux of the scattered C~{\sc iv} $\lambda$5800 emission present in the nebular spectrum (Figure~1) with the equivalent width of the stellar C~{\sc iv} $\lambda$5800 line, we estimate that scattered starlight accounts for only about 12\% of the observed continuum, and so its effect is small.

We also use the weak temperature dependence of some He~{\sc i} line ratios to derive  $T_{\rm e}$.  Table~\ref{tab:diagnostics} shows that the He~{\sc i} line ratios yield much lower temperatures (4.1kK-5.1kK) than those derived from the CEL diagnostics (11.1kK-15.2kK).  The observed relation $T_{\rm e}$([O~{\sc iii}]) $>$ $T_{\rm e}$(BJ) $>$ $T_{\rm e}$(He~{\sc i}) is what would be expected in the presence of hydrogen-deficient inclusions within the nebula (Liu et al. 2001b).

\begin{table*}
\caption{Plasma diagnostics}
\label{tab:diagnostics}
\centering
\begin{tabular}{llcc}
\hline
\hline
\noalign{\smallskip}
Ion& Lines & Ratio &Result\\
\noalign{\smallskip}
\hline
\noalign{\smallskip}
  & & &$\log N_{\rm e}$\,(cm$^{-3}$)$^{\mathrm{a}}$\\
{[{S}~{\sc ii}]}& $I(\lambda6731)/I(\lambda6716)$  &  1.10 & 2.96\\
{[{O}~{\sc ii}]}& $I(\lambda3729)/I(\lambda3726)$  & 0.83 & 3.00\\
{[{O}~{\sc ii}]}& $I(\lambda3727)/I(\lambda7325)$  &16.85 & 3.38 \\
{[{Cl}~{\sc iii}]}& $I(\lambda5537)/I(\lambda5517)$ & 0.95 & 3.22\\
{[{Ar}~{\sc iv}]}& $I(\lambda4740)/I(\lambda4711)$ & 0.75& 2.77\\
\noalign{\vskip5pt}
 & & &$T_{\rm e}$\,(K)$^{\mathrm{b}}$\\
{[{N}~{\sc ii}]}& $I(\lambda6548+\lambda6584)/I(\lambda5754)$ & 52.60 & 12600\\
{[{O}~{\sc iii}]} & $I(\lambda4959+\lambda5007)/I(\lambda4363)$ &148.7& 11100\\
He~{\sc i}&$I(\lambda5876)/I(\lambda4471) $ &3.03&4100\\
He~{\sc i}&$I(\lambda6678)/I(\lambda4471) $ &0.86&5100\\
He~{\sc i}&$I(\lambda6678)/I(\lambda7281) $ &5.88&4800\\
He~{\sc i}&$I(\lambda5876)/I(\lambda7281) $ &20.84&6500\\
 BJ/H\,11 &     & 0.125 & 9400 \\
          &     &       & $\pm$3000 \\
\noalign{\smallskip}
\hline
 \end{tabular}
 \begin{list}{}{}
 \item[$^{\mathrm{a}}$] Assuming $T_{\rm e}=11\,100$\,K;
 \item[$^{\mathrm{b}}$] Assuming $\log N_{\rm e}({\rm cm}^{-3})=3.0$.
 \end{list}
 \end{table*}

\subsubsection{Collisionally Excited Lines (CELs)}
 
The ionic abundances derived from optical CELs are listed in Table~\ref{tab:cel_ab}.  They were derived using a constant temperature of 11.1~kK and a density of 1000\,cm$^{-3}$.  As discussed in Section~\ref{sub:plasma}, recombination excitation may contribute significantly to the flux in the [O~{\sc ii}] $\lambda$3727,3729 lines, and so the O$^{+}$/H$^{+}$ abundance derived from them should be considered an upper limit.  As the fraction of O in the form of O$^{+}$ is very small ($\le$2 per cent), the error this introduces into the total oxygen abundance determination is negligible.

\begin{table}
\caption{\label{tab:cel_ab} Empirical ionic abundances of heavy elements derived from CELs.}
\centering
\begin{tabular}{llr}
\hline
\hline
\noalign{\smallskip}
X$^{i+}$/H$^{+}$ & Lines & $N_{{\rm X}^{i+}}/N_{{\rm H}^+}$\\
\noalign{\smallskip}
\hline
\noalign{\smallskip}
N$^+$       & [{N}~{\sc ii}] $\lambda\lambda$6548, 6584 & 1.30(-6)\\
\noalign{\smallskip}
O$^+$       & [{O}~{\sc ii}] $\lambda\lambda$3726, 3729 & 6.37(-6)\\
O$^{2+}$    & [{O}~{\sc iii}] $\lambda\lambda$4959, 5007 & 2.96(-4)\\
\noalign{\smallskip}
Ne$^{2+}$   & [{Ne}~{\sc iii}]$\lambda\lambda$3868, 3967  & 6.59(-5)\\
\noalign{\smallskip}
S$^+$       & [{S}~{\sc ii}]  $\lambda\lambda$6716, 6731& 7.85(-8)\\
S$^{2+}$    & [{S}~{\sc iii}] $\lambda6312$& 2.22(-6)\\
\noalign{\smallskip}
Cl$^{2+}$   & [{Cl}~{\sc iii}] $\lambda\lambda5517, 5537$& 7.00(-8)\\
Cl$^{3+}$   & [{Cl}~{\sc iv}] $\lambda\lambda7531, 8046$& 4.20(-8)\\
\noalign{\smallskip}
Ar$^{2+}$   & [{Ar}~{\sc iii}] $\lambda\lambda7135, 7751$ &9.47(-7)\\
Ar$^{3+}$   & [{Ar}~{\sc iv}]  $\lambda\lambda4711, 4740$ &3.09(-7)\\
\noalign{\smallskip}
Fe$^{2+}$   & [{Fe}~{\sc iii}] $\lambda$4658& 3.82(-8)\\
\noalign{\smallskip}
\hline
\end{tabular}
\end{table}

\begin{table}
\caption{\label{tab:orl_ab} Empirical ionic abundances for NGC\,1501 from ORLs. The abundances listed between square brackets were not used in the calculation of the averages. See text for discussion.}
\centering
\begin{tabular}{lllccc}
\hline
\hline
\noalign{\smallskip}
Ion &$\lambda$\,(\AA)& Mult& trans & $I$   & $N_{{\rm X}^{i+}}/N_{{\rm H}^+}$\\
\noalign{\smallskip}
\hline
He$^+$   &4471.50 & V14   & &3.758  &7.29(-2) \\                                      
         &5876.66 & V11   & &11.398 &7.32(-2) \\
         &6678.16 & V46   & &3.215  &7.22(-2) \\
         &Average &       & &       &7.29(-2) \\
He$^{2+}$&4685.68 & 3.4   & &40.299 &3.03(-2) \\
He/H     &        &       & &       &1.03(-1) \\
\noalign{\vskip8pt}                   
C$^{2+}$ &4267.15 & V6    & 3d-4f &0.987  &8.25(-4) \\
         &7231.32 & V3    & 3p-3d &0.131  &[3.02(-4)] \\
         &Adopted &       &       &       &8.25(-4) \\
C$^{3+}$ &4186.90 & V18   & 4f-5g &0.276  &2.56(-3) \\
         &4651.47 & V1    & 3s-3p &1.462  &[7.00(-3)] \\
         &Adopted &       &       &       &2.56(-3) \\
C$^{3+}$ &4658.30 &       & 5-6   &0.100  &1.36(-4) \\
\noalign{\vskip8pt}          
N$^{2+}$ &4035.08 & V39a  &  &0.192  &1.83(-3) \\
         &4621.39 & V5    &  &0.100  &3.29(-3) \\
         &5666.63 & V3    &  &0.124  &1.02(-3) \\
         &5676.02 & V3    &  &0.154  &2.85(-3) \\
         &Average &       &  &       &2.25(-3) \\
\noalign{\vskip8pt}                 
O$^{2+}$ &3882.29 &V11,V12& 3p-3d & 0.369  &[8.90(-3)] \\
         &4285.69 & V78b  & 3d-4f & 0.186  &7.75(-3) \\
         &4307.23 & V53b  & 3d-4f & 0.151  &1.14(-2) \\
         &4357.25 & V63a  & 3d-4f & 0.108  &[1.48(-2)] \\
         &4477.90 & V88   & 3d-4f & 0.114  &1.05(-2) \\
         &4610.20 & V92c  & 3d-4f & 0.133  &7.75(-3) \\
         &4661.63 & V1    & 3s-3p & 0.451  &[3.52(-3)] \\
         &4696.35 & V1    & 3s-3p & 0.138  &[1.16(-2)] \\
         &Average &       &       &       & 9.35(-3) \\
\noalign{\vskip8pt}                  
Ne$^{2+}$&3777.14 & V1    & &0.278   &2.20(-3) \\
         &4397.99 & V57b  & &0.109  &[3.26(-3)] \\
         &4409.30 & V55e  & &0.113  &[1.76(-3)] \\
         &4413.22 & V65   & &0.157  &[6.80(-3)] \\
         &4430.94 & V61a  & &0.136  &[4.98(-3)] \\
         &4457.05 & V61d  & &0.124  &[1.30(-3)] \\
         &Average &       & &       &2.20(-3) \\
\noalign{\vskip8pt}         
Mg$^{2+}$&4481.21 & V4    & &0.064  &6.67(-5) \\
\hline
\end{tabular}
\end{table}

\subsubsection{Optical Recombination Lines (ORLs)}
\label{sub:orl_obs}

The ionic abundances derived from the measured ORL spectrum of NGC~1501 are listed in Table~\ref{tab:orl_ab}.  The ORL abundances were derived using the mean of the measured He~{\sc i} temperature of 5.1~kK, which is expected to be more representative of the temperature in the cold knots that give rise to most of the ORL flux rather than the [O~{\sc iii}] temperature or the BJ temperature.  The weak dependence of ORL abundances on physical conditions means that the difference between abundances derived adopting the three temperatures are not large.

The He$^{+}$/H$^{+}$ abundance was derived from the intensities of the $\lambda$4471, $\lambda$5876 and $\lambda$6678 He~{\sc i} lines, weighted 1:3:1 according to their approximate intensity ratios.  According to the formulae given in \citet{kingdon95}, collisional effects are negligible at the temperature of 5.1~kK adopted here.  The He$^{2+}$/H$^{+}$ abundance was derived from the He~{\sc ii} $\lambda$4686 line.

Heavy element ORLs are well detected in our spectrum.  Our adopted ORL ionic abundances for C, N and O are the mean of the abundances derived from the selected individual lines for each ion. In Table~\ref{tab:orl_ab} the ionic abundances that appear between square brackets were left out of the final averages for reasons that will be discussed next. The C~{\sc ii}~$\lambda$5342 line is clearly too strong and must be contaminated by a blend. For this reason, this line is not listed in Table~\ref{tab:orl_ab}. The V3 C~{\sc ii}~$\lambda$7231 line is a component of a 3p-3d transition, less reliable than the 3d-4f  C~{\sc ii}~$\lambda$4267 transition, and is therefore excluded from the average. For C$^{3+}$ we have preferred to use only the abundance derived from the V18 line at 4186.90~{\AA}, as this is a 4f-5g transition, hence more reliable than the V1 line at 4651.47~{\AA}, which is a 3s-3p transition. Only one C~{\sc iv} line is detected in our spectrum and this is blended with [Fe~{\sc iii}]~$\lambda$4658; we attributed 45\% of the total emission to the C~{\sc iv} 5-6 transition (see discussion in Section~3.7), obtaining C$^{4+}$/H$^+$~=~1.36\,$\cdot$\,10$^{-4}$ by number.
For O$^{2+}$ only the abundances derived from 3d-4f transitions were finally used for the average, with the exception of the V63a component at 4357.25~{\AA} which gives a high O$^{2+}$ abundance and may be affected by blending (Tsamis et al. 2004, found that this line is 2.7 times too strong in the spectrum of NGC~3242). In the case of Ne$^{2+}$, as previously noted by \citet{liu00}, the 3d-4f  coefficients are derived assuming the fine-structure levels of the ground term 2p$^4$~$^3$P$_{2,1,0}$ of Ne$^{2+}$ to be thermalised. However, given that the mean  $N_{\rm e}$ derived for NGC~1501 is more than an order of magnitude lower than the critical densities of the $^3$P$_1$ (2.0$\cdot$10$^5$ cm$^{-3}$) and  $^3$P$_0$ (2.9$\cdot$10$^{4}$ cm$^{-3}$) levels of Ne$^{2+}$, the assumption of Boltzmann equilibrium is probably unrealistic. At the lower densities of NGC~1501 we expect the $^3$P$_1$ and $^3$P$_0$ levels to be underpopulated and the $^3$P$_2$ level to be overpopulated. As the strongest 3d-4f lines originate from the $^3$P$_2$ level, their recombination coefficient would be underestimated leading to correspondingly overestimated abundances. From our observations the mean abundance derived from 3d-4f transitions is a factor 1.7 higher than the abundance yielded by the 3s-3p V1 multiplet. In the light of the above discussion we adopt the Ne$^{2+}$/H$^+$ abundance derived from Ne~{\sc ii} multiplet {\sc vi} only. 

\begin{table}
  \caption{\label{tab:icfs} ICFs and total elemental abundances by number with respect to H derived from CELs and ORLs.}
\centering
\begin{tabular}{lcccc}
\hline
\hline
\noalign{\smallskip}
Ion   & \multicolumn{2}{c}{ICF} & \multicolumn{2}{c}{Abundance} \\
      &  KB94   &  MC2        &  KB94   &  MC2                  \\
\noalign{\smallskip}
\hline
\multicolumn{5}{l}{from CELs}\\
N     &    60.4   & 252  & 7.85(-5) & 3.28(-4) \\
O     &    1.27   & 1.13 & 3.85(-4) & 3.42(-4) \\ 
Ne    &    1.30   & 1.06 & 8.57(-5) & 6.98(-5) \\
S     &    2.73   & 14.4 & 6.29(-6) & 3.31(-5) \\
Cl    &    2.83   & 1.13 & 1.98(-7) & 1.27(-7) \\
Ar    &    1.02   & 1.03 & 1.28(-6) & 1.29(-6) \\
Fe    &     --    & 155  &  --      & 5.92(-6) \\
   \multicolumn{5}{l}{from ORLs}               \\
C     &    1.02   &      & 3.52(-3) &          \\
N     &    1.02   &      & 2.29(-3) &          \\
O     &    1.30   &      & 1.22(-2) &          \\ 
Ne    &    1.30   &      & 2.86(-3) &          \\
\noalign{\smallskip}
\hline
\end{tabular}
\\
\end{table}

\subsection{Summary of Nebular Empirical Abundances}
\label{sub:empiricaldiscussion}
Total elemental abundances were first of all derived using the ionisation correction factor scheme of \citet[][KB94]{kingsburgh94} (see columns~2 and~4 of Table~\ref{tab:icfs}).  For oxygen, O$^{+}$/H$^{+}$ and O$^{2+}$/H$^{+}$ abundances are available from CELs, and the unseen O$^{3+}$ was corrected for using

\begin{equation}
\frac{\rm O}{\rm H} = \left(\frac{{\rm He}^{+} + {\rm He}^{2+}}{{\rm He}^{+}}\right)^{2/3} \cdot \left(\frac{{\rm O}^{+}+{\rm O}^{2+}}{{\rm H}^{+}}\right).
\end{equation}

The ionisation correction factor derived from this is 1.27.  For oxygen ORLs, only O$^{2+}$/H$^{+}$ is available, and to derive a total abundance it was assumed that the ORL O$^{+}$/O$^{2+}$ ratio was the same as the value of 0.022 derived from CELs, giving a total ICF of 1.30. It is worth noting at this point that the ionic fractions in the ORL- and CEL-emitting regions may actually differ, but, while we cannot constrain the exact amount, we expect that only a small fraction of oxygen to be in the form of O$^+$, even in the coldest regions \citep[e.g.][]{ercolano03III}, and so the errors in the determination of O/H should be small.

The nitrogen abundance from CELs is rather uncertain -- only [N~{\sc ii}] lines are seen, and N$^{+}$ accounts for only a small fraction of the total nitrogen abundance. KB94 give N/H = (N$^{+}$/H$^{+}$)$\cdot$(O/O$^{+}$), which gives an ionisation correction factor of 60.39. Uncertainties of the empirical determination of the ICF for N are discussed toward the end of this section. In the case of the ORLs, N$^{2+}$ and N$^{3+}$ abundances are available. The unseen N$^{+}$ is corrected for by assuming that N/N$^{+}$=O/O$^{+}$, which results in an ionisation correction factor of 1.02.

No carbon CELs are seen as only optical nebular spectra were available. As discussed in Section~\ref{sub:IUE}, the IUE observations of this object are unsuitable for nebular analysis. However, nebular C~{\sc ii}, C~{\sc iii} and C~{\sc iv} ORLs are seen in our optical spectra, and a total carbon ORL abundance of 3.52\,$\cdot$10$^{-3}$ by number with respect to hydrogen is derived.

One magnesium line is seen in our spectrum -- that of Mg~{\sc ii}~$\lambda$4481. No ionisation correction scheme exists for magnesium, but Mg$^{2+}$ occupies a very large ionisation potential interval, from $\sim$15\,eV to $\sim$80\,eV, and so almost all the magnesium present will be in the form of Mg$^{2+}$. The abundance derived for Mg$^{2+}$ is noteworthy.  While all the second-row elements have ORL abundances which are much higher than those derived from CELs and exceed the solar abundance by factors of 10-30, the abundance derived for magnesium, using the model MC2's ICF (see Table~\ref{tab:icfs} and Secions~4 and 5), is only a factor of 1.9 higher than the solar value.  This is in line with the Mg abundances found by \citet{barlow03} for other nebulae showing large CNO ORL/CEL discrepancies, and strongly suggests that the large CNO ORL abundances are due to astrophysical effects such as H-deficient knots, and not to some unknown atomic process, since the latter might be expected to affect third-row as well as second-row elements. 

Argon, sulphur and chlorine abundances are derived from CELs only (see Table~\ref{tab:cel_ab}), with KB94 ionisation correction factors of 1.02, 2.73 and 2.83, respectively.

The [Fe~{\sc iii}]~$\lambda$4658 line is contaminated by the C~{\sc iv}~$\lambda$4658 
line (see Table~\ref{tab:linelist}). However, from the measured intensity of [Fe~{\sc iii}]~$\lambda$4881 (Table~1), the second strongest line in the optical spectrum of [Fe~{\sc iii}], and the theoretical [Fe~{\sc iii}]~$\lambda$4658/$\lambda$4881 intensity ratio of 2.0 at $T_{\rm e}$\,=12~kK and N$_{\rm e}$\,=\,10$^3$\,cm$^{-3}$ \citep[collision strength and A-value from][respectively]{zhang96, nahar96}, we estimate that [Fe~{\sc iii}] contributes 55\% to the 4658~{\AA} blend, with C~{\sc iv} contributing the rest. This yields Fe$^{2+}$/H$^+$~=~3.82$\cdot$10$^{-8}$ (see Table~\ref{tab:cel_ab}). Using an ICF of 155 from the model MC2, we derive a CEL-based total Fe abundance of 5.92\,$\cdot$\,10$^{-6}$, by number with respect to hydrogen. 

We also calculated total abundances from the CELs using the ICFs from model MC2. Columns~3 and~5, in the upper half of Table~\ref{tab:icfs}, list the model ICFs and corresponding total abundances, obtained by applying the ICFs to the ionic abundances derived from the CEL analysis. Total elemental abundances obtained using the MC2 ICFs should be more accurate than those obtained by using the KB94 ICFs and are adopted in this work. The largest discrepancy between the KB94 ICFs and the model ones is found for nitrogen. 
The CEL ionic abundances, together with the KB94 ICFs, give N/O~=~0.20, which would make NGC~1501 a non-Type~I PN according to both the criterion of \citet[][N/O~$>$~0.5 for a Type~I PN]{peimbert83} and the criterion of KB94 (N/O~$>$~0.85 for a Milky Way Type~I PN). However, the factor of four difference between the KB94 and MC2 ICFs for nitrogen turns NGC~1501 into a Type~I PN when using the MC2 ICFs, with (N/O)$_{\rm CEL}$~=~0.96. 

For the ORLs, when using MC2 ICFs we find N/O~=~0.52, while for the KB94 ICFs we find N/O~=~0.19. However, since CEL abundances are believed to provide a better representation of `average' nebular gas (see e.g. P\'equignot et al. 2003), the classification of NGC~1501 as a Type~I PN is preferred here. The elemental abundances derived from the CELs using MC2 ICFs are listed in Table~\ref{tab:abco}, where they are compared with the ORL-based abundances, with average local disk PN abundances from KB94, and with solar abundances. For N, O and Ne, the ORL-based abundances are larger than the CEL-based abundances by factors of 7, 36 and 42, respectively.

\begin{table}
\caption{Recommended elemental abundances in NGC\,1501 from CELs and ORLs, in units such that ${\rm log}\,N(H)=12.0$}
\label{tab:abco}
\centering
\begin{tabular}{ccccc}
\hline
\hline
\noalign{\smallskip}
Element& ORLs & CELs   & PN$^{\mathrm{a}}$ & Solar$^{\mathrm{b}}$\\
   &      &        & Average &     \\
\noalign{\smallskip}
\hline
\noalign{\smallskip}
He & 11.01 &   --   & 11.06   & 10.93\\
C  & 9.55  &   --   & 8.74    &  8.39\\
N  & 9.36  & 8.52   & 8.35    &  7.92\\
O  & 10.09 & 8.53   & 8.68    &  8.69\\
Ne & 9.46  & 7.84   & 8.09    &  8.08\\
S  &  --   & 7.52   & 6.92    &  7.33\\
Cl &  --   & 5.10   &  --     &  5.50\\
Ar &  --   & 6.11   & 6.39    &  6.40\\
Fe &  --   & 6.77   &  --     &  7.52\\
\noalign{\smallskip}
\hline
\end{tabular}
\begin{list}{}{}
 \item[$^{\mathrm{a}}$] The overall average abundances of Galactic PNe (Kingsburgh \& Barlow, 1994);
 \item[$^{\mathrm{b}}$] Solar abundances from Grevesse \& Sauval (1998, 1999), apart from C and O abundances which are from \citet{allende01, allende02}. 
\end{list}
\end{table}

\section{Three-dimensional Photoionisation Modelling of the Nebula}

\subsection{The Computer Code and Model Setup}
The three-dimensional photoionisation code \mbox{{\sc mocassin}}--Version~1.10 \citep{ercolano03I} was used in order to construct a realistic model for NGC\,1501. Sets of models were run, of which we selected two, MC1 and MC2, which best reproduced the nebular emission spectrum. The initial models (MC1) were run on the HiPerSPACE facilities at University College London, consisting of a SUN Microsystems V880 multiprocessor computer with 16 processors and 32Gb of memory. The later models (MC2) were run on a Beowulf cluster that belongs to the Institute of Computational Mathematics and Scientific/Engineering Computing of the Chinese Academy of Sciences, consisting of 120 550-MHz Pentium~{\sc iii} processors with 1Gb of memory each. A grid of 71$^3$ spatial points was used for our final models, corresponding to 357911 cubic cells of length 1405~AU each (assuming a distance of 1300~pc). The only difference between the MC1 and MC2 models is that the latter uses test values (see Section~5.1.1) for the unknown low-temperature dielectronic recombination coefficients of third-row elements (S, Cl and Ar), while the former sets all unavailable rates to zero. The nebular and stellar input parameters, as well as the final model abundances for MC1 and MC2, are listed in Table~\ref{tab:mocassininput}.  

\begin{table}
\begin{center}
\caption{Input parameters for the \mbox{{\sc mocassin}} photoionisation models. The abundances listed in this table are the final abundances used in models MC1 and MC2}
\begin{tabular}{lc|ccc}
\hline
\hline
\noalign{\smallskip}
\multicolumn{2}{c}{Stellar and Nebular} & & \multicolumn{2}{c}{Nebular Abundances} \\
\multicolumn{2}{c}{ Parameters}  &       &  Model & CEL \\
\hline
$T_{\rm eff}$ [kK]          & 110   & He/H  & 0.110    &  - \\
log\,$g$ (cgs)              & 6.0   & C/H   & 3.40(-4) &  - \\
$L_{\rm *}$ [L$_{\odot}$]   & 5000  & N/H   & 3.16(-4) & 3.28(-4) \\
$D$ [kpc]                   & 1.3   & O/H   & 3.50(-4) & 3.42(-4) \\
$\epsilon$                  & 0.423 & Ne/H  & 6.35(-5) & 6.98(-5) \\
$R_{\rm cube}$ [\arcsec]    & 38    & Mg/H  & 3.80(-5) &  - \\
                            &       & Si/H  & 1.00(-5) &  - \\
                            &       & S/H   & 3.10(-5) & 3.31(-5) \\
                            &       & Cl/H  & 1.20(-7) & 1.27(-7) \\
                            &       & Ar/H  & 1.10(-6) & 1.29(-6) \\
                            &       & Fe/H  & 6.60(-6) & 5.92(-6) \\

\hline
\label{tab:mocassininput}
\end{tabular}
\end{center}
\end{table}

\subsection{The Density Distribution}

The three-dimensional density distribution grid used for the modelling was determined for by Ragazzoni et al. (2001) from long-slit echellograms, reduced according to the methodology developed by \citet{sabbadin00a,sabbadin00b}. As the density distribution is, in general, a difficult input parameter to constrain, given the plurality of acceptable solutions, the existence, in this case, of a model density directly derived from the observations puts us in a favourable position, allowing a potentially higher degree of accuracy in the determination of the other nebular and central star parameters. The data cube made available to us by F.\,Sabbadin consisted of a 200$^3$ grid of  $N_{\rm e}$; these were mapped to the 71$^3$ \mbox{{\sc mocassin}} grid and converted to gas densities by iteratively using the local ionisation structure calculated at each grid cell to calculate the conversion factors. \\

Figures~4 and~5 of \citet{ragazzoni01} show iso-density plots of the  $N_{\rm e}$ distribution they derived at two different threshold values ($N_{\rm e}$~=~300\,cm$^{-3}$ and $N_{\rm e}$~=~900\,cm$^{-3}$), as seen from 12 different directions, while the optical appearance of the rebuilt nebula as seen from the same 12 directions is shown in their Figure~6. It is worth noting at this point that the original  $N_{\rm e}$ cube contained values of $N_{\rm e}$(H$\alpha$), where $N_{\rm e}$(H$\alpha$)~=~$N_{\rm e}$(S~{\sc ii})\,$\cdot$\,$\epsilon^{0.5}$, $\epsilon$ being the local filling factor. Unfortunately, the value of $\epsilon$ could not be determined from Ragazzoni et al.'s observations. The value of 0.423 used in the \mbox{{\sc mocassin}} models (and reported in Table~\ref{tab:mocassininput}) was chosen in order to reproduce $I$(H$\beta$)~=~5.26\,10$^{-11}$\,erg\,s$^{-1}$\,cm$^{-2}$, the integrated H($\beta$) flux from Cahn et al. (1992), dereddened using c(H$\beta$)\,=\,1.0 (see Section~\ref{sub:reddening}). This was made possible by the fact that this large, old, low-electron density PN is optically thin in H~{\sc i}, implying that the H$\beta$ emission of the nebula is not dependent on the luminosity of the central star, but represents a measure of the ionised mass. From this it is clear that I(H$\alpha$)\,$\propto$\,$\frac{N_{\rm e}^2\epsilon}{D^2}$, where $D$ is the distance from the nebula. As is true for many astronomical objects, the distance to NGC\,1501 is quite poorly constrained. At least 15 estimates are available in the literature, including statistical and individual distance estimates, with values ranging from 0.9 kpc \citep{amnuel84} to 2.0\,kpc \citep{acker78}. The value of 1.30\,kpc adopted in this work, was chosen for reasons of consistency with the value adopted by Ragazzoni et al. (2001) for their calculation of the three-dimensional ionisation structure from observations. We finally note that, although S$^+$ may not be a faithful representation of the ionised gas in this high-excitation nebula, Ragazzoni et al.'s result is certainly an improvement from the assumption of a spherically symmetric analytical density law.

\subsection{Elemental Abundances}
A homogeneous elemental abundance distribution was used for our models given that, although the ORL-CEL ADFs are high in this nebula, which may indicate the presence of cold ionised hydrogen-deficient material \citep[e.g.][]{liu00}, we do not have any direct observational evidence for the presence of structures containing plasma with different elemental abundances. It is, however, possible that such structures might exist inside the nebular shell, but with sizes falling below the spatial resolution of the instruments currently available. With the HST, a spatial resolution of 0.045\arcsec can be achieved in the optical -- this corresponds to 53~AU at 1300~pc. HST/WFPC2 images of NGC\,1501 have been taken as part of GO program 6119. However, the observations were aimed at detecting companions to central stars, and the short exposures necessary to obtain good stellar images yield very poor signal-to-noise ratios for the nebula itself. Also, the observations were taken in broad-band filters F555W and F814W, which contain many emission lines and are thus impossible to use for quantitative nebular analysis. As will be discussed in more detail in Section~\ref{sec:knots}, the presence of small, hydrogen-deficient knots could provide an explanation for the aforementioned ORL-CEL discrepancies as well as other spectroscopic peculiarities. 

The abundances derived empirically from CELs (see Section~\ref{sub:plasma}) were used as starting values for our modelling; these were then modified in order to obtain a better fit to the CEL spectrum. 

\subsection{Central Star Parameters}
\label{sub:cspar}
The effective temperature and gravity of the central star were varied in order to provide an ionising spectrum that could reproduce the ionisation structure implied by the observed nebular spectrum, under the constraints of our adopted density distribution, filling factor and distance. The final values adopted are $T_{\rm eff}$\,=\,110kK and log\,$g$\,=\,6.0 (cgs). This is well within the range of effective temperature determinations available in the literature, which range from $T_{\rm Zanstra}$(He~{\sc ii})~=~81kK \citep{stanghellini94} to the value of 135kK obtained by \citet{koesteke97} by means of stellar atmosphere modelling. It is worth noting at this point that, within a certain range of gravities, the value of log\,$g$ has a negligible influence on the ionising flux level of these models. 

 The hydrogen-deficient non-LTE stellar atmospheres used in this work were calculated using the T\"ubingen NLTE Model Atmosphere Package \citep{rauch03}, for He:C::N:O = 33:50:2:15 by mass, i.e. abundances close to those derived for the central star of NGC~1501 and for PG~1159 stars (see Section 3.4). These models are available for download from Dr. T. Rauch's website ({\tt http://astro.uni-tuebingen.de/\raisebox{1mm}{$\sim$}rauch}). The best results for the nebular photoionisation modelling were obtained with a central star effective temperature of 110~kK (as mentioned above) and a stellar luminosity of $L_{\rm *}$~=~5000\,L$_{\odot}$, which is close to the 5500\,L$_{\odot}$ estimate of \citet{koesteke97}. This model was found to be consistent with the level of the stellar UV continuum between 1300-1800~\AA\ in the dereddened IUE spectrum (see Section~2.2).

\subsection{The Ionising Spectrum}

As discussed in Section~\ref{sub:cs1501}, evidence for a strong stellar wind is observed in the spectrum of the central star of NGC~1501. This poses the question of whether a model atmosphere calculated with the plane parallel T\"ubingen code is a justifiable choice. We compared it with a model atmosphere calculated with the spherically symmetric code CMFgen \citep{hillier98}, for wind parameters close to those of the central star of NGC~1501 and abundances He:C:N:O~=~54:36:1:8 by mass, similar to those derived by Koesteke \& Hamman (1997; He:C:O~=~50:35:15), with a T\"ubingen model for the PG1159-type abundances given above; both models had $T_{\rm eff}$\,=\,135kK. Despite the different abundances, the variations in the flux levels between the two models were found to be less than a factor of two,  for wavelengths shortward of the Lyman limit, which is within the error bars imposed by the uncertainties of the stellar abundance determination. We also compared our T\"ubingen model atmosphere at 110kK with one calculated by L. Koesteke (private communication) using a spherically symmetric wind model for the same temperature and PG1159-type abundances, and found that the flux levels predicted by the two models agreed to within 20\%.


\section{Nebular Modelling: Strategy and Results}

The results of our photoionisation modelling and a comparison with observed values are presented in this Section. Our preliminary modelling was aimed at constraining the basic stellar and nebular parameters. Bearing in mind that the empirical analysis of the ORL spectrum of NGC~1501 yields larger abundances than the CEL analysis, and that this could be interpreted as indicating the existence of a separate component in the nebular gas, our efforts aimed to match the CEL spectrum only, as this should be representative of the dominant mass component of the nebular gas \citep{liu00,pequignot03}. 

\subsection{The first model (MC1)}

The initial model will be referred to in this paper as MC1. The CEL intensities predicted by MC1 are compared to the observed values in Table~\ref{tab:celmodel}. Table~\ref{tab:moddiagnostics} lists the predicted and observed nebular diagnostic line ratios. The results obtained from this first round of modelling, of which only the best-fitting model is shown in the table, were quite surprising. A comparison of the CEL intensities predicted by the model with the observations indicates that the model is too highly ionised. However this is at odds with what is implied by the predicted He~{\sc i} and He~{\sc ii} recombination lines, which show that the degree of ionisation of He is actually slightly lower than indicated by observations (see Table~13). The problem is evident from examination of the values listed in Table~\ref{tab:ionicratio}, containing the X$^{i+1}$/X$^{+i}$ model ionic fractions (columns~2 and 3) and those derived from the empirical analysis (column~4). Changes of the central star ionising spectrum, aimed to adjust the ionisation degree of O, S, Cl and Ar in the nebula, would necessarily also affect the ionisation structure of He, thus causing a larger discrepancy with the observations. This was confirmed by a comparison of models run with several $T_{\rm eff}$-log~$g$ combinations within the range $T_{\rm eff}$~=~100kK-140kK and log~$g~=~$5.0-8.0~[cgs]. 


 \begin{table}
 \centering
 \caption{Collisionally excited emission line intensities predicted by \mbox{{\sc mocassin}}, compared to observed values. The line intensities are given relative to $I$(H$\beta$) on a scale where $I$(H$\beta$)\,=\,100. The intensity of H$\beta$ is given in absolute units [erg~cm$^{-2}$~s$^{-1}$] and is compared to the value given by \citep{cahn92}, dereddened by c(H$\beta$)\,=\,1.0. }
 \label{tab:celmodel}
 \begin{tabular}{lccc}
 \hline
 \hline
 \noalign{\smallskip}
 Line                                           &{\sc MC1} & {\sc MC2} & Observed \\
 \noalign{\smallskip}
 \hline
 \noalign{\smallskip}
  Log[I(H$\beta$)]                              & -10.27   & -10.27   & -10.28 \\
  \noalign{\smallskip}
  H$\beta$~$\lambda$4861		        & 100      & 100       & 100    \\
  \noalign{\smallskip}
  $[$N~{\sc ii}$]$~$\lambda$5755		& 0.20     & 0.20      & 0.25  \\ 
  $[$N~{\sc ii}$]$~$\lambda$6548		& 3.44     & 3.45      & 3.16  \\
  $[$N~{\sc ii}$]$~$\lambda$6584		& 10.14    & 10.01     & 9.94  \\	
  \noalign{\smallskip}
  $[$O~{\sc ii}$]$~$\lambda$3727		& 5.44     & 10.7      & 12.16 \\
  $[$O~{\sc ii}$]$~$\lambda$3729		& 4.69     & 9.23      & 10.08 \\
  $[$O~{\sc ii}$]$~$\lambda\lambda$7318,9	& 0.21     & 0.41      & 0.73  \\
  $[$O~{\sc ii}$]$~$\lambda\lambda$7330,0	& 0.17     & 0.33      & 0.59  \\
  \noalign{\smallskip}
  $[$O~{\sc iii}$]$~$\lambda$4363	        & 12.81    & 12.78     & 10.38 \\
  $[$O~{\sc iii}$]$~$\lambda$4932	        & 0.19     & 0.19      & 0.35  \\
  $[$O~{\sc iii}$]$~$\lambda$4959	        & 458      & 456       & 394   \\
  $[$O~{\sc iii}$]$~$\lambda$5007	        & 1319     & 1312      & 1151  \\
  \noalign{\smallskip}
  $[$Ne~{\sc iii}$]$~$\lambda$3869	        & 102      & 103       & 101   \\
  $[$Ne~{\sc iii}$]$~$\lambda$3967	        & 31.6     & 31.7      & 28.10 \\
  \noalign{\smallskip}
  $[$S~{\sc ii}$]$~$\lambda$4069		& 0.069    & 0.28      & 1.49  \\
  $[$S~{\sc ii}$]$~$\lambda$4076		& 0.023    & 0.10     & 0.32 \\
  $[$S~{\sc ii}$]$~$\lambda$6717		& 0.43     & 1.76      & 1.77  \\
  $[$S~{\sc ii}$]$~$\lambda$6731		& 0.48     & 1.98      & 1.95  \\
  \noalign{\smallskip}
  $[$S~{\sc iii}$]$~$\lambda$6312$^*$	        & 1.48     & 1.48      & 1.48 \\
  \noalign{\smallskip}
  $[$Cl~{\sc iii}$]$~$\lambda$5518   	        & 0.11     & 0.84      & 0.76 \\
  $[$Cl~{\sc iii}$]$~$\lambda$5538   	        & 0.090    & 0.71      & 0.72 \\
  \noalign{\smallskip}
  $[$Cl~{\sc iv}$]$~$\lambda$7531    	        & 0.56     & 0.26      & 0.26 \\
  $[$Cl~{\sc iv}$]$~$\lambda$8046    	        & 1.31     & 0.61      & 0.54 \\
  \noalign{\smallskip}
  $[$Ar~{\sc iii}$]$~$\lambda$7136   	        & 1.30     & 13.08     & 14.34 \\
  $[$Ar~{\sc iii}$]$~$\lambda$7752   	        & 0.31     & 3.16      & 3.52  \\
  \noalign{\smallskip}
  $[$Ar~{\sc iv}$]$~$\lambda$4712    	        & 9.02     & 2.74      & 2.66 \\
  $[$Ar~{\sc iv}$]$~$\lambda$4741    	        & 7.01     & 2.12      & 2.00 \\
  \noalign{\smallskip}
 \hline
 \end{tabular}
\\
\small{$^*$ Blend with He~{\sc ii}~$\lambda$6310}
\end{table}

\begin{table*}
\caption{Calculated and observed plasma diagnostic ratios}
\label{tab:moddiagnostics}
\centering
\begin{tabular}{llcccc}
\hline
\hline
\noalign{\smallskip}
Ion& Lines                                                      & IP\,(eV)&  Observed & MC1   & MC2   \\
\noalign{\smallskip}
\hline
\noalign{\smallskip}
{[{S}~{\sc ii}]}& $I(\lambda6731)/I(\lambda6716)$               & 10.36   & 1.10      & 1.12  & 1.12  \\
{[{O}~{\sc ii}]}& $I(\lambda3729)/I(\lambda3726)$               & 13.62   & 0.83      & 0.86  & 0.86  \\
{[{O}~{\sc ii}]}& $I(\lambda3727)/I(\lambda7325)$               & 13.62   & 16.85     & 26.84 & 26.11 \\
{[{Cl}~{\sc iii}]}& $I(\lambda5537)/I(\lambda5517)$             & 23.81   & 0.95      & 0.84  & 0.84  \\
{[{Ar}~{\sc iv}]}& $I(\lambda4740)/I(\lambda4711)$              & 40.74   & 0.75      & 0.78  & 0.77  \\
\noalign{\vskip5pt}
{[{N}~{\sc ii}]}& $I(\lambda6548+\lambda6584)/I(\lambda5754)$   & 14.53   & 52.38     & 68.67 & 68.60 \\
{[{O}~{\sc iii}]} & $I(\lambda4959+\lambda5007)/I(\lambda4363)$ & 35.12   & 148.7     & 138.7 & 138.4 \\
He~{\sc i}&$I(\lambda5876)/I(\lambda4471) $                     & 24.59   & 3.03      & 2.72  & 2.72  \\
He~{\sc i}&$I(\lambda6678)/I(\lambda4471) $                     & 24.59   & 0.86      & 0.76  & 0.76  \\
\noalign{\smallskip}
\hline
 \end{tabular}
 \end{table*}

 \begin{table}
 \tabcolsep 10pt
 \centering
 \caption{\label{tab:hemodel} Helium recombination line intensities predicted by {\sc MOCASSIN}, compared to observed values. The line intensities are given relative to $I$(H$\beta$) on a scale where $I$(H$\beta$)\,=\,100. The intensity of H$\beta$ is given in absolute units [erg~cm$^{-2}$~s$^{-1}$] and is compared to the value given by \citep{cahn92}, dereddened by c(H$\beta$)\,=\,1.0. }
 \label{tab:helines}
 \begin{tabular}{lccc}
 \hline
 \hline
 \noalign{\smallskip}
 Line                           & {\sc MC1} & {\sc MC2} & Observed \\
 \noalign{\smallskip}
 \hline
 \noalign{\smallskip}
 Log[I(H$\beta$)]               &  -10.27   & -10.27    & -10.28\\
\noalign{\smallskip}
 H$\beta$~$\lambda$4861	        & 100       & 100       & 100   \\
\noalign{\smallskip}
 He~{\sc i}~$\lambda$4471       & 4.20      & 4.20      & 3.76 \\
 He~{\sc i}~$\lambda$4922       & 1.11      & 1.11      & 1.07 \\
 He~{\sc i}~$\lambda$5876       & 11.42     & 11.43     & 11.40 \\
 He~{\sc  i}~$\lambda$6678      & 3.21      & 3.21      & 3.12  \\ 
 He~{\sc i}~$\lambda$7065       & 2.16      & 2.16      & 2.51 \\
\noalign{\smallskip}
 He~{\sc ii}~$\lambda$4686      & 34.7      & 34.6      & 40.30 \\
\noalign{\smallskip}
 \noalign{\smallskip}
 \hline
 \end{tabular}
 \end{table}

\begin{table}
\caption{Predicted and observed ionic ratios}
\label{tab:ionicratio}
\centering
\begin{tabular}{lccccc}
\hline
\hline
\noalign{\smallskip}
X$^{i+1}/$X$^i$  &  MC1 & MC2& obs & MC1/obs  & MC2/obs\\
\hline
O$^{2+}$/O$^+$      & 116   & 58.9  & 46.5  & 2.49 & 1.27  \\
S$^{2+}$/S$^+$      & 46.8  & 27.8  & 28.3  & 1.65 & 0.98  \\
Cl$^{3+}$/Cl$^{2+}$ & 10.4  & 0.61  & 0.60  & 17.4 & 1.02  \\
Ar$^{3+}$/Ar$^{2+}$ & 12.0  & 0.35  & 0.33  & 36.2 & 1.06  \\
\noalign{\smallskip}
\hline
\hline
X$^{i+1}/$X$^i$  &  MC1 & MC2 & obs & MC1/obs & MC2/obs  \\
\hline
He$^{2+}$/He$^+$    & 0.36 & 0.36 & 0.41   & 0.88 & 0.88   \\
\noalign{\smallskip}
\hline
\end{tabular}
\end{table}

\subsubsection{The problem of data incompleteness for the low-temperature dielectronic recombination process}
\begin{table}
\caption{Estimated upper limits to low-temperature dielectronic rates}
\label{tab:rates}
\centering
\begin{tabular}{lcc}
\hline
\hline
\noalign{\smallskip}
X$^{i}$  & f(X$^i$)$^a$ &  $\alpha_D(X^{i},10^4K)$ \\
         &              &   [cm$^3$\,s$^{-1}$]     \\
\hline
S$^{+}$   & 3.14        & 5.66(-12)                \\
Cl$^{2+}$ & 17.5        & 5.81(-11)                \\
Ar$^{2+}$ & 42.5        & 13.7(-11)                \\
\noalign{\smallskip}
\hline
\end{tabular}
\\
\small{$^a$f(X$^i$)~=~$\alpha_D(X^{i},10^4K)$/$\alpha_R(X^{i},10^4K)$}
\end{table}

\begin{table}
\caption{Low-temperature dielectronic rates for CNO ions \citep{nussbaumerstorey84}}
\label{tab:CNOrates}
\centering
\begin{tabular}{lcc}
\hline
\hline
\noalign{\smallskip}
X$^{i}$ & f(X$^i$)$^a$ &  $\alpha_D(X^{i},10^4K)$      \\
        &          &  [cm$^3$\,s$^{-1}$]         \\
\hline
C$^{+}$   & 2.4    & 6.06(-12) \\
C$^{2+}$  & 2.7    & 1.36(-11) \\
N$^{+}$   & .86    & 2.03(-12) \\
N$^{2+}$  & 4.1    & 2.16(-11) \\
O$^{+}$   & .83    & 1.66(-12) \\
O$^{2+}$  & 2.0    & 1.14(-11) \\
\noalign{\smallskip}
\hline
\end{tabular}
\\
\small{$^a$f(X$^i$)~=~$\alpha_D(X^{i},10^4K)$/$\alpha_R(X^{i},10^4K)$}
\end{table}

The over-ionisation problem of S, Cl, and Ar could be interpreted as evidence for the effects of low-temperature dielectronic recombination, a process which can in magnitude be comparable (or often larger) than its radiative counterpart, but for which no reliable rates have yet been calculated for third-row elements. The uncertainty of the atomic data available for some astrophysically abundant elements is faced by all photoionisation codes. \citet{ali91} argued that, for a given ion X$^{+i}$, a better estimate than zero is obtained by taking the average of the rate coefficients for C$^{+i}$, O$^{+i}$ and N$^{+i}$, given that the coefficients seem to follow a certain trend with ionisation stage. \citet{dudziak00}, use coefficients empirically calibrated from a new unpublished model of the PN~NGC~7027. In this work, we do not attempt to calculate empirical coefficients, as NGC~1501, with its complicated nebular spectrum and WR central star, cannot be considered a typical PN.  Self-consistent calculations of dielectronic and radiative rates of some third-row elements are currently being carried out (Storey, Ercolano and Badnell, in preparation), but in this paper we will limit ourselves to seeking empirical upper limits to the unknown dielectronic rates and use these coefficients as test values for our nebular modelling. 

For a pocket of gas in ionisation equilibrium, the abundance ratio of two successive ionic stages is given by 
\begin{equation}
\frac{X^{i+1}}{X^i} \sim \frac{N_{\rm phot}^{X^{\rm +i}}}{N_{\rm e}\cdot(\alpha_R(X^{\rm +i},T_{\rm e})+\alpha_{\rm D}(X^{\rm +i},T_{\rm e}))}
\end{equation}
where $N_{\rm phot}^{X^{\rm +i}}$ is the total number of photoionisation events and $\alpha_R(X^{\rm +i},T_e)$ and $\alpha_D(X^{+i},T_e)$ are the temperature-dependent rate coefficients for, respectively, radiative and dielectronic recombination to X$^{+i}$. Note that we are assuming that the contribution due to charge exchange processes is negligible in this case, given that the nebula is optically thin in H~{\sc i}. This implies that we can use the ratio of model to observed X$^{i+1}$/{X$^i$} values to obtain an estimate of $\alpha_D(X^{+i},T_{\rm e})$: 
\begin{equation}
\alpha_D(X^{+i},T_e) = \left(  \frac{ (X^{i+1}/X^{+i})^{mod}}{ (X^{i+1}/X^{+i})^{obs}} - 1 \right)\,\cdot\,\alpha_R(X^{+i},T_e)
\label{eq:alphad}
\end{equation}
Unfortunately, our observations did not yield all the {X$^{i+1}$}/{X$^{+i}$} ratios necessary to derive a consistent set of estimates for all the relevant ionic stages of S, Cl and Ar, but only for S$^+$, Ar$^{2+}$ and Cl$^{2+}$. We define a quantity $f(X^{+i})$ such that the dielectronic rates for each ionic stage can be expressed as $\alpha_D$(X$^{+i}$,10$^4$K)~=~$f(X^{+i})$\,$\alpha_R$(X$^{+i}$,10$^4$K), where initial guesses for $f(X^{+i})$ were taken from Equation~\ref{eq:alphad} and then adjusted iteratively with the modelling. From the above, $f(X^{+i})$ is defined as the ratio of the low-temperature dielectronic recombination coefficient to the radiative recombination coefficient of the ion $X^{\rm i}$ at $T_{\rm e}$~=~10$^4$K. The use of a temperature independent coefficient such as $f(X^{+i})$ is equivalent to saying that the dielectronic rates follow the same temperature dependence as the radiative rates for the same ion, which is, of course, not true. Nevertheless, this is the best we can do in this case, as we are not attempting to calculate precise rates, a task that can only be fulfilled by laboratory measurements or detailed atomic modelling of each individual ion.  Furthermore, the nebular gas  $T_{\rm e}$'s do not show large deviations from a mean of 11100~K, and therefore the error introduced by the temperature dependence will be much smaller than the margin of error implied by our empirical approach. The final $f(X^{+i})$ values used are given in Table~\ref{tab:rates}, together with the corresponding values of $\alpha_D(X^{+i},T_{\rm e})$ for $T_{\rm e}$~=~10$^4$K. These values are only to be considered as upper limits, since other effects may  be at play in complicating the ionisation structure of this object (see Section~\ref{sec:knots}). As a comparison, the low-temperature dielectronic recombination coefficients and the corresponding $f$ values are listed in Table~\ref{tab:CNOrates} for C, N and O ions \citep{nussbaumerstorey84} at 10kK. Using the Ali et al. (1991) method method described above with the values listed in Table~\ref{tab:CNOrates}, one obtains estimates of 3.25\,$\cdot$\,10$^{-12}$cm$^3$\,s$^{-1}$ and 1.55\,$\cdot$\,10$^{-11}$cm$^3$\,s$^{-1}$ for the singly- and doubly-ionised cases at $T_{\rm e}$\,=\,10kK, lower than the values listed in Table~\ref{tab:rates}, necessary to resolve the discrepancy with the observations.

\subsection{Models including estimated low-temperature dielectronic recombination rates for third-row elements (MC2)}

\subsubsection{The CEL spectrum}

The CEL intensities predicted by the best-fitting model from a second set of models (MC2), which use the estimated upper limits to the dielectronic recombination rates given in Table~\ref{tab:rates}, are given in column~3 of Table~\ref{tab:celmodel}. This model provides a much better fit to the overall observed emission line spectrum than model MC1, but some problems are still present, as discussed in Section~6. 

The ionisation degree of O also appeared to be overestimated by model MC1 (see Table~\ref{tab:ionicratio}). The dielectronic rates included in the atomic database of \mbox{{\sc mocassin}}, were obtained by \citet{nussbaumerstorey84} who performed the calculations within the LS coupling scheme. As discussed in Section~4.3 of the Nussbaumer \& Storey (1984) paper, using intermediate coupling (IC) \citep[e.g.][] {beigman80} could result in higher rates as more states are taken into account (although not all of these would be permitted to autoionise). With this in mind, we increased the dielectronic rates in MC2 by a factor of three, which returned a better fit of the oxygen CEL spectrum. However, as discussed in Section~6, the large corrections to the atomic data required by MC2 are questionable and highlight the limitations of this chemically homogeneous model. 

The [O~{\sc ii}]\,$\lambda$7319 and $\lambda$7330 doublets and the [S~{\sc ii}]\,$\lambda$4069 and $\lambda$4076 lines are still underestimated by model MC2. Recombination processes can contribute to the intensity of these lines, and, whilst effective recombination coefficients to the metastable levels of [O~{\sc ii}] are known \citep{liu00}, the same is not true for [S~{\sc ii}]. We have included the recombination term to the statistical equilibrium matrix in our photionisation model, in order to self-consistently calculate the populations of the metastable levels of [O~{\sc ii}], by taking into account recombination and collisional population and depopulation. However, at the temperatures and densities of the nebular gas, we found recombination contributions to be negligible. We therefore conclude that the discrepancy between model and observed intensities of these line must be attributed to recombination processes occurring in a separate, colder and/or denser gas phase (see Section~6).  

\subsubsection{The computed ionic and thermal structure of the nebula}
\label{sub:tandi}

\renewcommand{\baselinestretch}{1.2}
\begin{table*}
\begin{center}
\caption{Nebular-averaged fractional ionic abundances, obtained from model MC2.}
\begin{tabular}{lccccccc}
\multicolumn{8}{c}{} \\
\hline
  	&	&	&	& Ion	 &	&	&	\\
\cline{2-8}
Element	& {\sc i}&{\sc ii}&{\sc iii}&{\sc iv}&{\sc v}&{\sc vi}&{\sc vii}\\
\hline
H	& 1.27(-3) & 9.99(-1) &          & 	    & 	       &          &          \\
He	& 7.43(-4) & 7.34(-1) & 2.65(-1) &  	    &          &          &          \\
C	& 4.60(-6) & 8.89(-3) & 4.09(-1) & 5.67(-1) & 1.49(-2) &          &          \\
N	& 9.95(-7) & 3.97(-3) & 3.91(-1) & 6.02(-1) & 3.11(-3) &          &          \\
O	& 1.24(-5) & 1.47(-2) & 8.66(-1) & 1.18(-1) & 1.31(-3) &          &          \\
Ne	& 9.37(-7) & 5.03(-3) & 9.42(-1) & 5.30(-2) & 3.36(-5) &          &          \\
Mg      & 2.98(-3) & 3.11(-2) & 9.15(-1) & 5.12(-2) & 1.93(-6) &          &          \\
Si      & 5.40(-5) & 3.77(-2) & 2.64(-1) & 4.18(-1) & 2.80(-1) &          &          \\
S	& 3.36(-7) & 2.42(-3) & 6.72(-2) & 4.87(-1) & 4.39(-1) & 4.41(-3) & 5.89(-6) \\
Cl	& 1.13(-6) & 4.44(-3) & 5.51(-1) & 3.57(-1) & 1.05(-1) & 2.94(-3) & 1.12(-7) \\
Ar	& 1.18(-7) & 9.40(-4) & 7.14(-1) & 2.53(-1) & 2.90(-2) & 2.63(-3) & 9.37(-6) \\	
Fe	& 4.05(-5) & 6.72(-4) & 6.44(-3) & 6.47(-1) & 3.13(-1) & 3.22(-2) & 2.15(-5) \\
\hline
\label{tab:ionratio}
\end{tabular}
\end{center}
\end{table*}
\renewcommand{\baselinestretch}{1.5}

The nebular-averaged fractional ionic abundances calculated from the MC2 model are listed in Table~\ref{tab:ionratio}. Hydrogen and helium are both fully ionised and we see that significant fractions of the heavy elements are in higher ionisation stages. We notice that the N/N$^+$ ratio is nearly a factor of three larger than the O/O$^+$ ratio, contrary to what is generally assumed in ionisation correction schemes, including the one used in this work, where the two ionic fractions are taken to be the same. This has already been noticed by \citet{rubin88} and should be taken into account when considering uncertainties in empirical abundance analyses for optically thin PNe such as NGC~1501 (see also Section~3.7). 

We did not find large temperature fluctuations in the computed 3D temperature distribution, obtaining $t^2$(O$^{2+}$)~=~0.0021 from the model. This is not surprising, given that large density fluctuations are not present in our density distribution grid.

\section{The usual suspects: \lowercase{$t^2$} versus enhanced-density, metal-rich knots}
\label{sec:knots}

The introduction of the test values for the low-temperature dielectronic recombination coefficients of S, Ar and Cl certainly helped to achieve a better fit to the observed CEL spectrum of NGC~1501. However a number of problems still remain unsolved. In particular:
\begin{description}
\item[{1. The ORL-CEL discrepancy.}] MC2 fails to reproduce the observed ORL spectrum. The abundances derived empirically from the ORL spectrum are consistently larger than those derived from the CEL analysis, while the $T_{\rm e}$ derived from the observed He~{\sc i} recombination lines is cooler than the temperature estimates obtained from CELs (Table~5).
\item[{2. The large dielectronic rate value for argon.}] With regards to the Ar$^{3+}$/Ar$^{2+}$ ionic ratios, large low-temperature dielectronic recombination rates are required in order to resolve the discrepancy between the observations and the photoionisation models. Although strongly dependent on the atomic structure of each individual ion, atomic data currently available for a number of doubly ionised ions show low-temperature dielectronic rates to be at most ten times larger than the corresponding radiative rates (with average values being in the region of 3 or 4 times the radiative rates). On the other hand, recent work by \citet{morisset04} also pointed out the need for large dielectronic rates for Ar$^{2+}$ in order to reconcile ISO infrared observations and models for forty H~{\sc ii} regions in their sample.  
\end{description}

Temperature fluctuations have been proposed within the $t^2$ framework of \citet[][ and later references]{peimbert67,peimbert71} by \citet[e.g.][]{ruiz03} and \citet{peimbert04} as a solution to the ORL-CEL abundance discrepancy. In the case of NGC~1501 our three-dimensional models showed no evidence for $t^2$ fluctuations larger than 0.012. We also calculated empirical values for $T_{\rm 0}$([O~{\sc iii}]) and $t^2$([O~{\sc iii}]), following the formulations given by \citet{peimbert67,peimbert71} and, more recently \citet{peimbert04}. First of all we calculated that a value of $T_{\rm 0}$([O~{\sc iii}])~=~3750~K is required to obtain the same abundances from the [O~{\sc iii}] CELs and the O~{\sc ii} ORLs, implying 43\% temperature fluctuations ($t^2$([O~{\sc iii}])~=~0.184). However, when we derived values for the same quantities using the Balmer jump and the [O~{\sc iii}] temperatures from Table~\ref{tab:diagnostics}, we found a mean temperature of $T_{\rm 0}$([O~{\sc iii}])~=~10010~K and $t^2$([O~{\sc iii}])~=~0.036, which correspond to temperature fluctuations of only 19\%. We also calculated $T_{\rm 0}$ and $t^2$ values using the  $T_{\rm e}$ derived from the He~{\sc i} recombination lines (see Table~\ref{tab:diagnostics}). The He~{\sc i}~$\lambda$5876/$\lambda$4471  $T_{\rm e}$ of 4100~K implies that $T_{\rm 0}$~=~5350~K and $t^2$~=~0.154, while the mean He~{\sc i}  $T_{\rm e}$ of 5125~K implies that $T_{\rm 0}$~=~6300~K and $t^2$~=~0.133. These values of $T_{\rm 0}$ and $t^2$ would increase the O$^{2+}$ abundance measured from CELs by factors of 10.6, 3.3 and 6.5, respectively, still a long way short of resolving the factor of 30 discrepancy. The observational results and the computed 3D temperature structure indicate that something other than temperature fluctuations is causing the large ORL-CEL ADFs we found for NGC~1501.

The existence of metal-rich knots could  provide a solution to the problems mentioned above. While such knots have been invoked recently to help with the solution of the ORL-CEL discrepancy problem in PNe (e.g. Liu, 2000), and bi-abundance photoionisation models have been constructed for some PNe \citep{pequignot03}, it is further proposed here that the lower ionisation degree of heavy elements compared to helium shown by NGC~1501 can also be explained by the presence of small metal-rich knots mixed with the nebular gas. The knots could be formed by processed material transported out in the outflow from the AGB precursor of the H-deficient WR central star. 

Low-ionisation species would be abundant in the knots and in the shadow regions (tails) behind them, thanks to softening of the radiation field due to the screening effect of material that is optically thick to stellar UV radiation. However, due to the high metallicity and cooling rates in the material, the temperatures inside the knots are expected to be much too low for collisionally excited line radiation to be produced to a significant degree, while essentially all the ORL emission is expected to be produced in these regions \citep[see e.g. the 3-D photoionisation model of the H-poor knots in Abell~30 by ][]{ercolano03III}.

The situation in the shadowed tails of knots is completely different. Since these are composed of material with normal nebular abundances, they are not as efficiently cooled as the knots themselves, implying that the emission of collisionally excited lines from lower ionisation species would be expected. In this regard it is clear that the problem of over-ionisation of oxygen in the CEL-emitting region for models MC1 and MC2 could be eased by the introduction of such knots, which could also provide some of the missing emission from S$^+$, Ar$^{2+}$ and Cl$^{2+}$, with the implication that lower dielectronic rates than those listed in Table~15 would be sufficient to match the CEL spectrum. 

\section{Conclusions}

We have presented in this paper new, deep optical spectra of the PN NGC~1501. The empirical analysis led to the discovery of high ORL-CEL ADFs for  O$^{2+}$ and Ne$^{2+}$, with values of  32 and 33, respectively. The agreement between the ADFs for O and Ne lends further weight to the suggestion by \citet{liu00} that, while large discrepancies may exist between ORL and CEL abundances relative to H, elemental ratios such as Ne/O are often the same whether derived from CELs or ORLs. 

Unfortunately no temperature for the ORL-emitting region could be determined from our observations, since the O~{\sc ii}~$\lambda$4089 line \citep[see ][]{wesson03,tsamis04} was not detected in our observations. However, a large temperature discrepancy exists between the  $T_{\rm e}$ implied by the He~{\sc i} recombination line ratios (on average $\sim$5.1~kK) and the [O~{\sc iii}] temperature (11.4~kK). Given that we expect $T_{\rm e}({\rm ORL})~<~T_{\rm e}({\rm He}$~{\sc i})$~<~T_{\rm e}({\rm CEL})$ (Liu, 2003), the ORL emitting regions probably have temperatures lower than 4.5~kK, consistent with the results of \citet{wesson03} for knot J3 of Abell~30, where $T_{\rm e}({\rm ORL})$~=~2.5~kK. $T_{\rm e}({\rm He}$~{\sc i})$~>~T_{\rm e}({\rm ORL})$ is a logical consequence of He~{\sc i} recombination lines being produced in both the ORL and the CEL-emitting regions and, therefore, the  $T_{\rm e}$'s derived from them are not representative of either region. 

Our initial models showed metals to be too highly ionised compared to observations, except for helium. We suspect that two effects are at play in complicating the emission line spectrum of NGC~1501. First of all, we note the incompleteness of low-temperature dielectronic recombination coefficients for third-row elements and the possible underestimation of $\alpha_D$(O$^+$, $T$). We have obtained upper limits for $\alpha_D$(S$^+$, 10$^4$\,K), $\alpha_D$(Cl$^{2+}$, 10$^4$\,K) and $\alpha_D$(Ar$^{2+}$, 10$^4$\,K) of 5.66$\cdot$10$^{-12}$, 5.81$\cdot$10$^{-11}$ and 13.7$\cdot$10$^{-11}$~cm$^3$~s$^{-1}$, respectively. These correspond to low-temperature dielectronic recombination coefficients that are factors of 3.1, 17.5  and 42.5 higher than the corresponding radiative recombination rates for S$^{+}$,  Cl$^{2+}$ and Ar$^{2+}$, respectively. We also found that increasing $\alpha_D$(O$^+$, $T$) by a factor of three yields a better fit to the [O~{\sc ii}] spectrum. The very large rates required for Ar and Cl make it improbable that the discrepancies between the models and observations can be resolved solely in terms of these coefficients. This, together with the large ORL-CEL ADFs derived for NGC~1501, points to the existence of a second component, consisting of cold, metal-rich ORL-emitting knots. It seems that the solution to the over-ionisation problem may lie partly with the softening of the ionising radiation field, due to absorption of UV photons by such knots and partly with the lack/uncertainty of low-temperature dielectronic rates. Clearly the magnitudes of the dielectronic rates required to fit the spectrum could be reduced by an amount depending on the fraction of the emitting gas comprised by the knots, as well as their size and density, which will affect the size of the shadowed regions. 

We also note that NGC~1501 is yet another PN with a hydrogen-deficient central star which shows high ADFs. To our knowledge, NGC~1501, with its ADF of $\sim$32, has the third highest ADF of any nebula, after Abell~30 \citep[ADF$\sim$700;][]{wesson03} and Hf\,2-2 \citep[ADF$\sim$84;][]{liu03}. Hydrogen-deficient knots are directly observed in the former (Jacoby et al., 1979; Hazard et al., 1980) and three-dimensional photoionisation modelling (Ercolano et al., 2003b) has shown that in its case cold, metal-rich material can account for the ORL emission observed. In the case of Abell~30, a popular view with regards to the origins of such knots is the born-again scenario \citep{iben83}, whereby hydrogen-deficient material would have been stripped off the stellar surface during a late thermal pulse event.\\

There are, however, inconsistencies between the predictions of the born-again scenario and observations of some of the nebulae it applies to (including Abell~30). Most importantly, evolutionary models predict a late thermal pulse event to produce a carbon-rich stellar surface abundance \citep{herwig99}, whereas in the case of Abell~30 \citet{wesson03} measured (C/O)$_{\rm orl}$~=~0.8; similarly, for NGC~1501 we measured (C/O)$_{\rm orl}$~=~0.31. A reconciliation of theory and observations could be obtained if the ORLs were produced in a gas component consisting of carbon-rich material expelled from the stellar surface, which has already mixed somewhat with surrounding oxygen-rich nebular gas. In the case of a dense knot, one would expect the ionising radiation to be able to penetrate only a thin outer layer, the same boundary region would also be the most likely to be subject to mixing. If the above is true, one would then expect the measured nebular (C/O)$_{\rm orl}$ to be larger than (C/O)$_{\rm cel}$, which is indeed confirmed by the data available in the literature. In particular we looked at a ORL-CEL study of 16 PNe by \citet{tsamis04}, for which ten PNe had (C/O)$_{\rm orl}$/(C/O)$_{\rm cel}>$1, two lacked a value for (C/O)$_{\rm orl}$, two lacked a value for (C/O)$_{\rm cel}$, and only for two PNe was (C/O)$_{\rm orl}$/(C/O)$_{\rm cel}<$1. A similar trend is also confirmed by the work of Wesson (2004). \\
If the origin of the hydrogen-deficient knots can really be traced to material ejected from the surface of the AGB precursors of the central stars of born-again nebulae, one might also expect a correlation between hydrogen-deficient central stars and nebulae showing large ORL-CEL abundance discrepancies. Such a correlation did not emerge from investigation of the measured ORL-CEL ADFs \citep[][Wesson 2004]{liu95,liu00,liu01b,liu03,demarco01,tsamis04} and 
central star classification \citep[e.g.][]{demarco01} available in the literature. On the one hand, the central star spectral type is not always known and we cannot therefore exclude that some high-ADF central stars are not actually H-deficient.  For instance, the spectral type of the central star of the PN Hf~2-2 \citep[ADF=84; ][]{liu03} is unknown. On the other hand, the IUE spectra of the central star of PN DdDm~1 \citep[ADF=11.8][]{wesson04} has no emission lines of C~{\sc iv}~$\lambda$1550 or He~{\sc ii}~$\lambda$1640, which are expected in WR spectra. The jury is therefore still out as to whether the concept of H-deficient knots as an explanation for large ADFs can be reconciled with current post-AGB stellar evolutionary theories. We can however be certain that the high ADFs in nebulae such as NGC~1501 cannot be explained by nebular temperature fluctuations of the classically conceived type.\\

\vspace{7mm}
\noindent
{\bf Acknowledgements}

We are grateful to Franco Sabbadin for providing the three-dimensional density distribution file of NGC\,1501. We also thank Pete Storey, Christophe Morisset and Falk Herwig for helpful discussions. We thank Lars Koesteke for providing model atmospheres which account for NLTE, spherical expansion and metal line blanketing, and James Herald for the loan of the CMFgen model atmosphere used for comparison with the T\"ubingen model. We thank Bob Rubin for helpful comments on the revised version and the anonymous referee for a thorough report. We acknowledge the Institute of Computational Mathematics and Scientific/Engineering Computing of the Chinese Academy of Science for the use of the PC cluster of the Major State Basic Research Project (973 project) ``Large Scale Scientific Computation", LSSC-I. BE and RW acknowledge support from PPARC. OD acknowledges financial support from Janet Jeppson Asimov. TR is supported by the DLR under grant 50\,OR\,0201.

\bibliographystyle{mn2e}

\bibliography{references}

\begin{thebibliography}{}

\bibitem[\protect\citeauthoryear{Acker}{Acker}{1978}]{acker78}
Acker A.,  1978, A\&AS, 33, 367

\bibitem[\protect\citeauthoryear{Ali, Blum, Bumgardner, Cranmer, Ferland,
  Haefner \& Tiede}{Ali et~al.}{1991}]{ali91}
Ali B.,  Blum R.~D.,  Bumgardner T.~E.,  Cranmer S.~R.,  Ferland G.~J.,
  Haefner R.~I.,    Tiede G.~P.,  1991, PASP, 103, 1182

\bibitem[\protect\citeauthoryear{Allende~Prieto, Lambert \&
  Asplund}{Allende~Prieto et~al.}{2001}]{allende01}
Allende~Prieto C.,  Lambert D.~L.,    Asplund M.,  2001, ApJ, 556, L63

\bibitem[\protect\citeauthoryear{Allende~Prieto, Lambert \&
  Asplund}{Allende~Prieto et~al.}{2002}]{allende02}
Allende~Prieto C.,  Lambert D.~L.,    Asplund M.,  2002, ApJ, 573, L137

\bibitem[\protect\citeauthoryear{Aller}{Aller}{1976}]{aller76}
Aller L.~H.,  1976, MSRSL, 9, 271

\bibitem[\protect\citeauthoryear{Amnuel, Guseinov, Novruzova \&
  Rustamov}{Amnuel et~al.}{1984}]{amnuel84}
Amnuel P.~R.,  Guseinov O.~K.,  Novruzova K.~I.,    Rustamov I.~S.,  1984,
  Ap\&SS, 107, 19

\bibitem[\protect\citeauthoryear{Barlow \& Hummer}{Barlow \&
  Hummer}{1982}]{barlow82}
Barlow M.~J.,  Hummer D.~G.,  1982, in  de Loore, C. W. H., Willis, A. J., D. Reidel eds., 
IAU Symp Vol.~99, Wolf-Rayet stars: Observations, physics, evolution; Astr. Soc. Pac., San Francisco, p.~387

\bibitem[\protect\citeauthoryear{Barlow, Liu, P\'equignot, Storey, Tsamis \&
  C.}{Barlow et~al.}{2003}]{barlow03}
Barlow M.~J.,  Liu X.-W.,  P\'equignot D.,  Storey P.~J.,  Tsamis Y.~G.,    C.
  M.,  2003, in Planetary Nebulae: their evolution and role in the Universe; S.
  Kwok, in M. Dopita and R. Southerland eds., IAU Symp Vol.~209, Planetary Nebulae: 
their evolution and role in the Universe, Astr. Soc. Pac., San Francisco, p.~373

\bibitem[\protect\citeauthoryear{Beigman \& Chichkov}{Beigman \&
  Chichkov}{1980}]{beigman80}
Beigman I.~L.,  Chichkov B.~N.,  1980, JPhB, 13, 565

\bibitem[\protect\citeauthoryear{Bond \& Ciardullo}{Bond \&
  Ciardullo}{1993}]{bond93}
Bond H.~E.,  Ciardullo R.,  1993, in M. A. Barstow ed., NATO Advanced Science Institutes (ASI) Series C. Vol~403, 
White Dwarfs: Advances in Observation and  Theory, Kluwer, Dordrecht, p.~491

\bibitem[\protect\citeauthoryear{{Bond et al.}}{{Bond et al.}}{1996}]{bond96}
{Bond et al.} 1996, AJ, 112, 2699

\bibitem[\protect\citeauthoryear{Cahn, Kaler \& Stanghellini}{Cahn
  et~al.}{1992}]{cahn92}
Cahn J.~H.,  Kaler J.~B.,    Stanghellini L.,  1992, A\&AS, 94, 399

\bibitem[\protect\citeauthoryear{Ciardullo \& Bond}{Ciardullo \&
  Bond}{1996}]{ciardullo96}
Ciardullo R.,  Bond H.~E.,  1996, AJ, 111, 2332

\bibitem[\protect\citeauthoryear{Crowther, de Marco \& Barlow}{Crowther
  et~al.}{1998}]{crowther98}
Crowther P.~A.,  de Marco O.,    Barlow M.~J.,  1998, MNRAS, 296, 367

\bibitem[\protect\citeauthoryear{Crowther, Fullerton, Hillier, Brownsberger, K.
  Dessart, Willis, De~Marco, Barlow, Hutchings, J.~B. Massa, Morton \& D.~C.
  Sonneborn}{Crowther et~al.}{2000}]{crowther00}
Crowther P.~A.,  Fullerton A.~W.,  Hillier D.~J.,  Brownsberger K. Dessart L.,
  Willis A.~J.,  De~Marco O.,  Barlow M.~J.,  Hutchings J.~B. Massa D.~L.,
  Morton   D.~C. Sonneborn G.,  2000, ApJ, 538, L51

\bibitem[\protect\citeauthoryear{De~Marco \& Barlow}{De~Marco \&
  Barlow}{2001}]{demarco01}
De~Marco O.,  Barlow M.~J.,  2001, Ap\&SS, 275, 53

\bibitem[\protect\citeauthoryear{Drew, Barlow, Unruh, Parker, Wesson, Pierce,
  Mashder \& Phillipps}{Drew et~al.}{2004}]{drew04}
Drew J.~E.,  Barlow M.~J.,  Unruh Y.~C.,  Parker Q.~A.,  Wesson R.,  Pierce
  M.~J.,  Mashder M. R.~W.,    Phillipps S.,  2004, MNRAS, in press

\bibitem[\protect\citeauthoryear{Dudziak, P\'equignot, Zijlstra \&
  Walsh}{Dudziak et~al.}{2000}]{dudziak00}
Dudziak G.,  P\'equignot D.,  Zijlstra A.~A.,    Walsh J.~R.,  2000, A\&A, 363,
  717

\bibitem[\protect\citeauthoryear{Ercolano, Barlow, Storey \& Liu}{Ercolano
  et~al.}{003a}]{ercolano03I}
Ercolano B.,  Barlow M.~J.,  Storey P.~J.,    Liu X.-W.,  2003a, MNRAS, 340,
  1136

\bibitem[\protect\citeauthoryear{Ercolano, Barlow, Storey, Liu, Rauch \&
  Werner}{Ercolano et~al.}{003b}]{ercolano03III}
Ercolano B.,  Barlow M.~J.,  Storey P.~J.,  Liu X.-W.,  Rauch T.,    Werner K.,
   2003b, MNRAS, 344, 1145

\bibitem[\protect\citeauthoryear{Feibelman}{Feibelman}{1998}]{feibelman98}
Feibelman W.~A.,  1998, ApJS, 119, 197

\bibitem[\protect\citeauthoryear{Gorny \& Stasinska}{Gorny \&
  Stasinska}{1995}]{gorny95}
Gorny S.~K.,  Stasinska G.,  1995, A\&A, 303, 893

\bibitem[\protect\citeauthoryear{Herwig, Bloecker, Langer \& Driebe}{Herwig
  et~al.}{1999}]{herwig99}
Herwig F.,  Bloecker T.,  Langer N.,    Driebe T.,  1999, A\&A, 349, L5

\bibitem[\protect\citeauthoryear{Hillier \& Miller}{Hillier \&
  Miller}{1998}]{hillier98}
Hillier D.~J.,  Miller D.~L.,  1998, ApJ, 496, 407

\bibitem[\protect\citeauthoryear{Howarth}{Howarth}{1983}]{howarth83}
Howarth I.~D.,  1983, MNRAS, 203, 801

\bibitem[\protect\citeauthoryear{Iben \& Renzini}{Iben \&
  Renzini}{1983}]{iben83}
Iben I.~J.,  Renzini A.,  1983, ARA\&A, 21, 271

\bibitem[\protect\citeauthoryear{Kaler}{Kaler}{1976}]{kaler76}
Kaler J.~B.,  1976, ApJS, 31, 517

\bibitem[\protect\citeauthoryear{Kingdon \& Ferland}{Kingdon \&
  Ferland}{1995}]{kingdon95}
Kingdon J.~B.,  Ferland G.~J.,  1995, ApJ, 442, 714

\bibitem[\protect\citeauthoryear{Kingsburgh \& Barlow}{Kingsburgh \&
  Barlow}{1994}]{kingsburgh94}
Kingsburgh R.~L.,  Barlow M.~J.,  1994, MNRAS, 271, 257 (KB94)

\bibitem[\protect\citeauthoryear{Kingsburgh, Barlow \& Storey}{Kingsburgh
  et~al.}{1995}]{kingsburgh95}
Kingsburgh R.~L.,  Barlow M.~J.,    Storey P.~J.,  1995, A\&A, 295, 75 (KBS95)

\bibitem[\protect\citeauthoryear{Koesterke \& Hamann}{Koesterke \&
  Hamann}{1997}]{koesteke97}
Koesterke L.,  Hamann W.-R.,  1997, in H. J. Habing and H. J. G. L. M. Lamers eds., 
IAU ymp Vol.~180, Planetary nebulae, Kluwer, Dordrecht, p.~114

\bibitem[\protect\citeauthoryear{Liu}{Liu}{2002}]{liu02}
Liu X.-W.,  2002, in Ionized Gaseous Nebulae, RMxAA Vol.~12 of Eds. W. J.
  Henney, J. Franco, M. Martos, \& M. Pe\~na.
p.~70

\bibitem[\protect\citeauthoryear{Liu}{Liu}{2003}]{liu03}
Liu X.-W.,  2003, in Planetary Nebulae: their evolution and role in the Universe; S.
  Kwok, in M. Dopita and R. Southerland eds., IAU Symp Vol.~209, Planetary Nebulae: 
their evolution and role in the Universe, Astr. Soc. Pac., San Francisco, p.~373

\bibitem[\protect\citeauthoryear{Liu \& Danziger}{Liu \&
  Danziger}{1993}]{liu93}
Liu X.-W.,  Danziger I.~J.,  1993, MNRAS, 263, 256

\bibitem[\protect\citeauthoryear{Liu \& et}{Liu \& et}{001a}]{liu01a}
Liu X.-W.,  et a.,  2001a, MNRAS, 323, 343

\bibitem[\protect\citeauthoryear{Liu, Luo, Barlow, Danziger \& Storey}{Liu
  et~al.}{001b}]{liu01b}
Liu X.-W.,  Luo S.-G.,  Barlow M.~J.,  Danziger I.~J.,    Storey P.~J.,  2001b,
  MNRAS, 327, 141

\bibitem[\protect\citeauthoryear{Liu, Storey, Barlow \& Clegg}{Liu
  et~al.}{1995}]{liu95}
Liu X.-W.,  Storey P.~J.,  Barlow M.~J.,    Clegg R. E.~S.,  1995, MNRAS, 272,
  369

\bibitem[\protect\citeauthoryear{Liu, Storey, Barlow, Danziger, Cohen \&
  Bryce}{Liu et~al.}{2000}]{liu00}
Liu X.-W.,  Storey P.~J.,  Barlow M.~J.,  Danziger I.~J.,  Cohen M.,    Bryce
  M.,  2000, MNRAS, 312, 585

\bibitem[\protect\citeauthoryear{Morisset, Schaerer, Bouret \&
  Martins}{Morisset et~al.}{2004}]{morisset04}
Morisset C.,  Schaerer D.,  Bouret J.-C.,    Martins F.,  2004, A\&A, 415, 577

\bibitem[\protect\citeauthoryear{Nahar}{Nahar}{1996}]{nahar96}
Nahar S.~N.,  1996, A\&AS, 119, 509

\bibitem[\protect\citeauthoryear{Nussbaumer \& Storey}{Nussbaumer \&
  Storey}{1984}]{nussbaumerstorey84}
Nussbaumer H.,  Storey P.~J.,  1984, A\&AS, 56, 293

\bibitem[\protect\citeauthoryear{Peimbert}{Peimbert}{1967}]{peimbert67}
Peimbert M.,  1967, ApJ, 150, 825

\bibitem[\protect\citeauthoryear{Peimbert}{Peimbert}{1971}]{peimbert71}
Peimbert M.,  1971, Bol. Obs. Ton. Tac., 6, 29

\bibitem[\protect\citeauthoryear{Peimbert, Peimbert, Ruiz \& Esteban}{Peimbert
  et~al.}{2004}]{peimbert04}
Peimbert M.,  Peimbert A.,  Ruiz M.~T.,    Esteban C.,  2004, ApJS, 150, 431

\bibitem[\protect\citeauthoryear{Peimbert \& Torres-Peimbert}{Peimbert \&
  Torres-Peimbert}{1983}]{peimbert83}
Peimbert M.,  Torres-Peimbert S.,  1983, Planetary nebulae; Proceedings of the
  IAU Symposium, 103, 233

\bibitem[\protect\citeauthoryear{P\'equignot, W., J. \& C.}{P\'equignot
  et~al.}{2003}]{pequignot03}
P\'equignot D.,  Liu X.-W.,  Barlow M. J.,    Morisset C.,  2003,
 in M. Dopita and R. Southerland eds., IAU Symp Vol.~209, Planetary Nebulae: 
their evolution and role in the Universe, Astr. Soc. Pac., San Francisco, p.~347

\bibitem[\protect\citeauthoryear{Ragazzoni, Cappellaro, Benetti, Turatto \&
  Sabbadin}{Ragazzoni et~al.}{2001}]{ragazzoni01}
Ragazzoni R.,  Cappellaro E.,  Benetti S.,  Turatto M.,    Sabbadin F.,  2001,
  A\&A, 369, 1088

\bibitem[\protect\citeauthoryear{Rauch}{Rauch}{2003}]{rauch03}
Rauch T.,  2003, A\&A, 403, 709

\bibitem[\protect\citeauthoryear{Rubin}{Rubin}{1968}]{rubin97}
Rubin R.~H.,  1968, ApJ, 153, 761

\bibitem[\protect\citeauthoryear{Rubin}{Rubin}{1986}]{rubin86}
Rubin R.~H.,  1986, ApJ, 309, 334

\bibitem[\protect\citeauthoryear{Rubin, Simpson, Erickson \& Haas}{Rubin
  et~al.}{1988}]{rubin88}
Rubin R.~H.,  Simpson J.~P.,  Erickson E.~F.,    Haas M.~R.,  1988, ApJ, 327,
  377

\bibitem[\protect\citeauthoryear{Ruiz, Peimbert, Peimbert \& Esteban}{Ruiz
  et~al.}{2003}]{ruiz03}
Ruiz M.~T.,  Peimbert A.,  Peimbert M.,    Esteban C.,  2003, ApJ, 595, 247

\bibitem[\protect\citeauthoryear{Sabbadin, Benetti, Cappellaro \&
  Turatto}{Sabbadin et~al.}{000a}]{sabbadin00a}
Sabbadin F.,  Benetti S.,  Cappellaro E.,    Turatto M.,  2000a, A\&A, 361,
  1112

\bibitem[\protect\citeauthoryear{Sabbadin, Cappellaro, Benetti, Turatto \&
  Zanin}{Sabbadin et~al.}{000b}]{sabbadin00b}
Sabbadin F.,  Cappellaro E.,  Benetti S.,  Turatto M.,    Zanin C.,  2000b,
  A\&A, 355, 688

\bibitem[\protect\citeauthoryear{Sabbadin \& Hamzaoglu}{Sabbadin \&
  Hamzaoglu}{1982}]{sabbadin82}
Sabbadin F.,  Hamzaoglu E.,  1982, A\&AS, 50, 1

\bibitem[\protect\citeauthoryear{Stanghellini, Kaler \& Shaw}{Stanghellini
  et~al.}{1994}]{stanghellini94}
Stanghellini L.,  Kaler J.~B.,    Shaw R.~A.,  1994, A\&A, 291, 604

\bibitem[\protect\citeauthoryear{Storey \& Hummer}{Storey \&
  Hummer}{1995}]{storey95}
Storey P.~J.,  Hummer D.~G.,  1995, MNRAS, 272, 41

\bibitem[\protect\citeauthoryear{Tsamis, Barlow, Liu, Storey \&
  Danziger}{Tsamis et~al.}{2004}]{tsamis04}
Tsamis Y.~G.,  Barlow M.~J.,  Liu X.-W.,  Storey P.~J.,    Danziger I.~J.,
  2004, MNRAS, in press

\bibitem[\protect\citeauthoryear{Tylenda}{Tylenda}{2003}]{tylenda03}
Tylenda R.,  2003,  in M. Dopita and R. Southerland eds., IAU Symp Vol.~209, Planetary Nebulae: 
their evolution and role in the Universe, Astr. Soc. Pac., San Francisco, p.~389

\bibitem[\protect\citeauthoryear{Tylenda, Acker \& Stenholm}{Tylenda
  et~al.}{1993}]{tylenda93}
Tylenda R.,  Acker A.,    Stenholm B.,  1993, A\&AS, 102, 595

\bibitem[\protect\citeauthoryear{Werner, U. \& K.}{Werner
  et~al.}{1991}]{werner91}
Werner K., Heber U., Hunger K.,   1991, A\&A, 244, 437

\bibitem[\protect\citeauthoryear{Wesson}{Wesson}{2004}]{wesson04}
Wesson R.,  2004, PhD Thesis, University of London

\bibitem[\protect\citeauthoryear{Wesson, Liu \& Barlow}{Wesson
  et~al.}{2003}]{wesson03}
Wesson R.,  Liu X.-W.,    Barlow M.~J.,  2003, MNRAS, 340, 253

\bibitem[\protect\citeauthoryear{Zhang}{Zhang}{1996}]{zhang96}
Zhang H.~L.,  1996, A\&AS, 119, 509

\end{thebibliography}

\bsp
\end{document}